\documentclass[twocolumn]{aastex631}

\usepackage{amsmath}

\begin{document}

\title{Eruptivity Criteria for Solar Coronal Flux Ropes in Magnetohydrodynamic and Magnetofrictional Models}

\author{Oliver E. K. Rice}
\affiliation{Department of Mathematical Sciences, Durham University, Durham DH1 3LE, UK}

\author{Anthony R. Yeates}
\affiliation{Department of Mathematical Sciences, Durham University, Durham DH1 3LE, UK}

\accepted{13th August 2023}

\submitjournal{The Astrophysical Journal}

\begin{abstract}

We investigate which scalar quantity or quantities can best predict the loss of equilibrium and subsequent eruption of magnetic flux ropes in the solar corona. Our models are initialized with a potential magnetic arcade, which is then evolved by means of two effects on the lower boundary: firstly a gradual shearing of the arcade, modelling differential rotation on the solar surface, and secondly supergranular diffusion. These result in flux cancellation at the polarity inversion line and the formation of a twisted flux rope. We use three model setups: full magnetohydrodynamics (MHD) in cartesian coordinates, and the magnetofrictional model in both cartesian and polar coordinates. The flux ropes are translationally-invariant, allowing for very fast computational times and thus a comprehensive parameter study, comprising hundreds of simulations and thousands of eruptions. Similar flux rope behavior is observed using either magnetofriction or MHD, and there are several scalar criteria that could be used as proxies for eruptivity. The most consistent predictor of eruptions in either model is the squared current in the axial direction of the rope, normalised by the relative helicity, although a variation on the previously proposed ‘eruptivity index’ is also found to perform well in both the magnetofrictional and MHD simulations.

\end{abstract}

%\keywords{Solar Corona, Magnetic Fields, Eruptions, Helicity}

\section{Introduction} \label{sec:intro}

This paper builds on our previous work \citep{2022FrASS...9.9135R}, in which we studied the formation and eruption of magnetic flux ropes using a translationally-invariant (2.5D) cartesian magnetofrictional model. Here we expand the study to include two further models: a second magnetofrictional model in axisymmetric polar coordinates and a full magnetohydrodynamic (MHD) model in cartesian coordinates. Our aim is to verify that the results previously presented are valid both in magnetofriction and MHD, and that the qualitative behavior of flux ropes is similar in both cartesian and polar models.

Flux ropes are formed when a magnetic arcade in the solar corona is sheared and reconnects with itself at its base, forming a twisted bundle of magnetic flux \citep{2020RAA....20..165L}. These ropes can become unstable under certain conditions, releasing large amounts of energy in a coronal mass ejection \citep{2006SSRv..123..251F}. The mechanisms behind such eruptions and the conditions required to trigger them are not yet fully understood, despite being an active field of research for many decades. Approaches to the study of such eruptions and instabilities have ranged from simple analytic two-dimensional models \citep[e.g.][]{1974A&A....31..189K} to full 3D MHD simulations \citep[e.g.][]{2013ApJ...778...99L,2014ApJ...787...46L}, which have only become feasible relatively recently.

One approach to predicting eruptivity is to look for stability criteria motivated by simplified theoretical models - a good example is the torus instability \citep[e.g.][]{2006PhRvL..96y5002K}, which is particularly applicable where the flux rope itself forms a section of a circular structure. However, the availability of parametric numerical simulations of flux rope eruptions also opens up the possibility to look for measurable quantities after the fact that show a significant increase or decrease shortly before an eruption. This approach was taken by \citet{2017A&A...601A.125P}, who analysed the simulations of \citet{2013ApJ...778...99L,2014ApJ...787...46L}. Most scalar measurements of the system did not have any strong connection to eruptivity, but the so-called `eruptivity index' - the fraction of the relative magnetic helicity made up of the current-carrying component - exhibited a large increase in those simulations where the flux rope later erupted.

Based on a small number of MHD simulations \citep{2015ApJ...814..126Z}, \citet{2018ApJ...863...41Z} have also observed a consistent threshold in the eruptivity index above which the system becomes unstable. This result is notable as the evolution of the relative helicity itself differs significantly from the simulations of \citet{2013ApJ...778...99L}, whereas the eruptivity index itself behaves similarly in both studies. Motivated by these results, the eruptivity index has been estimated prior to the eruption of several structures in the real corona \citep[e.g.,][]{2021A&A...653A..69G}, and  found to be consistently high prior to eruptive events. Estimates of the index using observational data must necessarily extrapolate the coronal magnetic field from photospheric measurements, which adds a degree of uncertainty. This has been an active field of study for some time, and is usually accomplished using a force-free field extrapolation \citep[e.g.,][]{2012LRSP....9....5W}.

In a series of parametric simulations, \citet{2023A&A...669A..33P} analysed the evolution of the eruptivity index in a coronal jet, where the driven boundary conditions were applied for differing lengths of time. This study also found an increase in the eruptivity index prior to an eruption, although it is acknowledged that a larger parameter study is necessary to determine if there is a consistent threshold above which the system becomes unstable.

Our previous study on the eruptivity of flux ropes \citep{2022FrASS...9.9135R} sought a compromise between simple analytical models and full 3D MHD simulations, by using the magnetofrictional model in a 2.5D cartesian domain. In the magnetofrictional model, pioneered by \citet{1986ApJ...309..383Y}, the fluid equations are disregarded and instead replaced with a fictitious `velocity' field obtained explicitly from the magnetic field, while the MHD induction equation is retained. This model is computationally much faster than full MHD as it is not necessary to resolve waves within the fluid. Flux rope formation and eruptions are observed \citep[e.g.,][]{2006ApJ...641..577M,2020ApJS..250...28H}, as in full MHD. Morevover, 
\citet{1986ApJ...311..451C} propose that replacing the fluid equations in ideal MHD with a fictitious velocity proportional to the Lorentz force (as in magnetofriction) does not affect the linear stability properties of the system. Indeed, \citet{2013A&A...554A..77P} have shown that configurations taken from magnetofrictional simulations at the point of eruption still lead to an eruption in full MHD.

Our previous studyThis found that although the eruptivity index did usually increase before flux rope eruptions, there was not a consistent value at which an eruption would occur. We thus judged that this quantity was not a good predictor of eruptions. Instead, we found that (among others) the ratio of the rope current squared to the relative helicity satisfied the requirements of a good predictor, there being a threshold above which an eruption was very likely within a given number of days. We note that in this case the helicity and associated quantitites were calculated using a new 2.5D-specific definition, the use of which will be discussed further in this paper.

In this paper we improve upon our previous work by directly comparing results from the 2.5D magnetofrictional code to equivalent simulations in full MHD, using the LARE2D code \citep{2001JCoPh.171..151A}. Each simulation takes far longer than an equivalent using the magnetofrictional method, but we have nevertheless performed a large parameter study, varying (among other quantities) the plasma beta in place of the magnetofrictional relaxation rate. We have also undertaken an equivalent parameter study in polar coordinates, using a new axisymmetric magnetofrictional code. We compare the results from this model to those in the cartesian coordinate system, to test the validity of the general use of cartesian simulations when modelling the solar corona. 

By performing very similar studies using both MHD and magnetofriction in cartesian geometry, we can directly compare the two methods, evaluating which (if any) eruptivity criteria are consistently good predictors in both cases, and if so whether the thresholds for eruptivity are independent of the method used. The existence of such criteria would lend further credence to the use of the magnetofrictional method in place of full MHD when predicting flux rope eruptions with trigger mechanisms similar to those in our study.

In all three simulation setups we observe similar behavior to our original study, with flux rope formation and eruptions observed over a large range of parameters. In addition to flux ropes we also observe periodic `arcade eruptions' - rapid reconnection at the top of a sheared magnetic arcade (above the flux rope, if one is present). It is possible that these represent streamer blowouts or even `stealth CMEs' \citep{2012LRSP....9....3W}. However, we do not focus on these eruptions but rather the eruption of the flux rope itself. In general, as the photospheric diffusion rate increases, flux ropes form and erupt more quickly. In some of our simulations we observe that after an eruption a flux rope reforms and subsequent eruptions can take place. For each of the three simulation setups the parameter study comprises 320 runs, which has allowed us to observe more than 1000 flux rope eruptions.

We begin in Section \ref{sec:approach} by outlining the basis of the three models -- cartesian magnetofriction, axisymmetric polar magnetofriction and cartesian MHD. We then describe the diagnostic measurements of the system, including a discussion on the optimal definition of the relative helicity and related quantities. In Section \ref{sec:behavior}  we describe the structure and behavior of flux ropes in both MHD and magnetofriction, and their dependence on the photospheric flux cancellation rate $\eta_0$. Finally, in Section \ref{sec:skillscores} we discuss which diagnostic ratios are good predictors of eruptivity in all three models, and directly compare the results from the equivalent magnetofrictional and MHD models.

\section{Modelling Approach} \label{sec:approach}

\subsection{Equations and Numerical Implementation}

We use three models to model magnetic flux rope behavior. The first two use the magnetofrictional model \citep[e.g.][]{2012LRSP....9....6M} in either cartesian or polar coordinates, using our own code. The third is a full magetohydrodynamic (MHD) model in cartesian coordinates, using the LARE2D code \citep{2001JCoPh.171..151A}. The models are all translationally-invariant (2.5D) either in the $z$ direction in the cartesian simulations or in the longitudinal ($\phi$) direction in the polar simulations.

In the cartesian models the domain is a square box with coordinates $-1 < x < 1$, $0 < y < 1$ and $z$ the invariant direction, with equally-spaced grid cells. The polar simulations model an entire hemisphere of the corona, with coordinates $0 < \theta <\pi$ and $ 1 < r< 2.5$, and $\phi$ the invariant longitudinal direction. The grid cells here are evenly spaced in $s = \cos \theta$ and $\rho = \log r$ \citep{2014SoPh..289..631Y}, which increases the resolution in the areas of interest.

Both models directly employ Faraday's law, Ohm's law and Amp\'ere's law. However, whereas MHD accurately models the fluid dynamics by coupling these to Euler's fluid equations, in magnetofriction a fictitious velocity field is obtained explicitly from the magnetic field.

In dimensionless form, the MHD equations used in the LARE2D code are as follows:
\begin{eqnarray}
\frac{\partial \mathbf{B}}{\partial t}&=&-\nabla\times\mathbf{E}\\
\mathbf{E}&=&\eta \mathbf{j} - \mathbf{v}\times\mathbf{B} \\
\mu_0 \mathbf{j}&=& \nabla\times\mathbf{B}\\
\frac{\partial \rho}{\partial t}&=&- \nabla\cdot(\rho \mathbf{v})\\
\frac{\mathrm{D}\mathbf{v}}{\mathrm{D}t}&=&\frac{1}{\rho}\mathbf{j}\times\mathbf{B}
-\frac{1}{\rho}\nabla P + \mathbf{g}\\
\frac{\mathrm{D}\epsilon}{\mathrm{D}t}&=&-\frac{P} {\rho}\nabla\cdot\mathbf{v}+\frac
{\eta}{\rho}j^{2} \\
\epsilon &=& \frac{P}{\rho(\gamma-1)}  \label{eqn:gas},
\end{eqnarray}
where the variable quanitites are the magnetic field density $\textbf{B}$, the current density $\textbf{j}$, the plasma pressure $P$, the plasma density $\rho$ and the internal energy density $\epsilon$. The ratio of specific heats is taken to be $\gamma = 5/3$. We choose the gravitational field to be $\mathbf{g} = -\mathbf{e}_y/(y+1)^2$.  The constant $\eta$ represents the coronal magnetic diffusivity and $\mu_0$ is the permeability of free space, henceforth taken to be unity. For consistency between the three model setups we set the coronal diffusivity as $\eta = 5 \times 10^{-4}$ throughout. There is considerable variation in the literature as to this value \citep[e.g.][]{2010ApJ...708..314A,2006ApJ...641..577M}, 
so we choose it to be as small as possible whilst still being able to resolve current sheets at our chosen grid resolution. No explicit viscosity is used.

There is no fluid in the magnetofrictional model, and so the plasma density, pressure and internal energy are not considered. Instead, we replace the fluid equations with a closed expression for a fictitious fluid velocity $\mathbf{v}$. The full set of magnetofrictional equations are
\begin{eqnarray}
\frac{\partial \mathbf{B}}{\partial t}&=&-\nabla\times\mathbf{E}\\
\mathbf{E}&=&\eta \mathbf{j} - \mathbf{v}\times\mathbf{B} \\
\mu_0 \mathbf{j}&=& \nabla\times\mathbf{B}\\
\mathbf{v} &=& \nu_0\frac{(\nabla \times \mathbf{B}) \times \mathbf{B}}{B^2 + \delta e^{-\delta B^2}} + v_{\rm{out}}(y)\textbf{e}_y,
\end{eqnarray}
where we have introduced the constant magnetofrictional relaxation rate $\nu_0$, which determines the time it takes for the system to relax to a current-free state. The `softening' constant $\delta = 0.01$ is a small number which prevents the denominator becoming zero at magnetic null points. The $v_{\rm{out}}$ term models the effect of the solar wind \citep{1958ApJ...128..664P}, resulting in an additional upward velocity. 

In cartesian coordinates this outflow speed is taken to be
\begin{equation}
    v_{\rm{out}}(y) = v_1y ^{10} \mathbf{e}_y,
\end{equation}
in line with our previous work, and in polar coordinates we use a more realistic approximation to the Parker solar wind solution \citep{2021ApJ...923...57R}
\begin{equation}
    v_{\rm{out}}(r) = v_1\frac{r_1^2e^{-2r_c /r}}{r^2e^{-2r_c/r_1}}\mathbf{e}_r ,
\end{equation}
where the critical radius $r_c$ is 10 solar radii, $r_1=2.5$ solar radii is the upper extent of the domain and  $v_1$ is a constant wind speed factor, chosen to be 50 times the maximum shearing velocity. A realistic value is likely higher than this, but this compromise is made as the resultant high velocities at the top of the domain increase computation times significantly. Such an outflow term is not necessary in the MHD calculations as we instead impose a gravitational field on the fluid, which in the absence of a magnetic field would naturally reach an equilibrium with nonzero vertical velocity.

In our magnetofrictional code we ensure the solenoidal condition ($\nabla \cdot \mathbf{B} = 0$) to machine precision by describing the system in terms of a vector potential $\textbf{A}$, such that 
\begin{equation}
    \mathbf{B} = \nabla \times \textbf{A}.
\end{equation}
%One of the advantages of magnetofriction is that the entire state of the system can be described by this field.
The LARE2D code does not employ a vector potential but instead ensures the solenoidal condition using the constrained transport method of \citet{1988ApJ...332..659E}.

Numerically, in our magnetofrictional models the differential operators are calculated using Gauss' or Stokes' theorem as appropriate, and the variable fields are stored on a staggered grid \citep{1966ITAP...14..302Y}. In cartesian coordinates this is equivalent to using a central-difference scheme. The cartesian simulations are initialised with Python and run using Fortran 90, with a grid resolution of 256 x 128 cells. The axisymmetric polar simulations are run using Python with a grid resolution of 180 x 60 cells. 

We adopt dimensionless units throughout. In the axisymmetric polar model one distance unit is taken to be 1 solar radius. In the cartesian models one distance unit is around the width of the magnetic arcade, which varies considerably but is assumed to be around $35^\circ$ on the solar surface (or 0.44 solar radii), based on arcades observed in the axisymmetric simulations. Using the photospheric shearing rates (discussed in Section \ref{sec:bconds}) the time units can then be set as 27.4 days in the cartesian simulations and 19.5 days in the axisymmetric simulations. However, the significant differences in the two coordinate systems mean this should only be regarded as the correct order of magnitude and as such the timescales discussed later in the paper should only be treated as indicative.

\subsection{Initial Conditions}
\label{sec:initconds} 

\begin{figure*}[ht!]

\plottwo{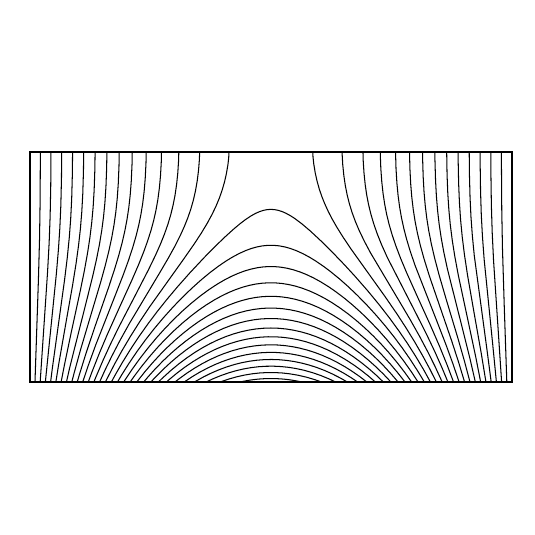}{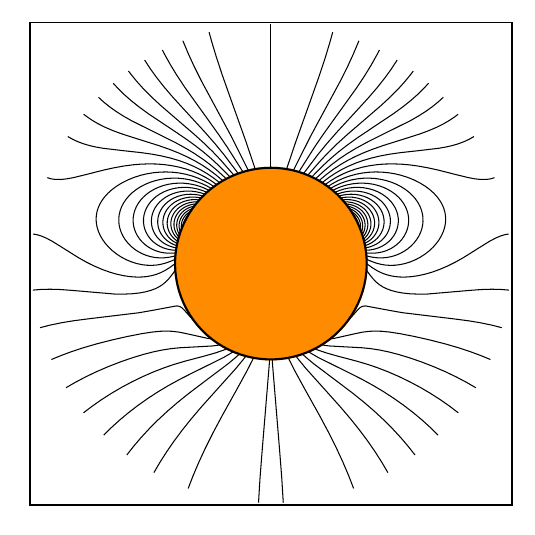}
\caption{Initial magnetic fields for the cartesian simulations (left) and the axisymmetric simulations (right), with magnetic field lines in black. These are PFSS fields with the lower radial boundary conditions as in Equations \eqref{eqn:initboundary} and  \eqref{eqn:initboundarycart}. In the axisymmetric simulations, the computational domain only covers one hemisphere, and the other is only plotted here for illustrative purposes.}
\label{fig:initial} 
\end{figure*}

The initial magnetic fields are plotted in Figure \ref{fig:initial}. In both cartesian and polar coordinates, the simulations are initialised with a PFSS (Potential Field Source-Surface) field, in which the magnetic field is radial at the upper boundary and there is no electric current throughout the domain. PFSS is chosen for consistency with the MHD simulations, although it would be preferable to use an `outflow field' \citep{2021ApJ...923...57R}, a steady-state field in which the magnetic field is in equilibrium when the solar wind is taken into account.

In polar coordinates, we choose the radial magnetic field strength on the lower boundary to fit the following analytic function:

\begin{equation}
    B_r(1,\theta) = s^7 + 5de^{-10d^2},  \label{eqn:initboundary}
\end{equation}
 where $s = \cos(\theta)$ and $d = s - \cos(0.35\pi)$. The first term ($s^7$) approximates the magnetic field of the global solar dipole reasonably well \citep[e.g.,][]{1978ARA&A..16..429S,2005ApJ...625..522W}, and the second term ensures there is a clear polarity inversion line at a latitude where the differential rotation of the surface will result in a considerable shearing rate. In the axisymmetric simulations we observe the arcade (the region with the closed field lines) between around 20 and 70 degrees from the north pole. A symmetric field around the equator would not be suitable as the differential rotation is symmetric between the northern and southern hemispheres, so such an arcade would merely be dragged around the equator without being sheared.

As in our previous work \citep{2022FrASS...9.9135R} the cartesian simulations (both magnetofriction and MHD) are initialised with the lower boundary condition 
\begin{equation}
    B_y(0,x) = B_0 \sin \left( \frac{\pi}{2}x \right ),  \label{eqn:initboundarycart}
\end{equation}
for some constant $B_0$ of order unity. 

The initial magnetic field completely specifies the magnetofrictional system, but in MHD we additionally require initial conditions for the fluid. The initial density is chosen to be constant throughout the domain, as is the initial internal energy. The initial fluid velocity is required to be positive in the $y$ direction in order to accurately emulate the solar wind -- otherwise it is possible for the fluid to be in equilibrium with a negative vertical velocity. The initial velocity profile chosen is $V_{\rm{out}}(y) = v_1 y^2$ for some constant $v_1$.

\begin{figure*}[ht!]
\centering
\plotone{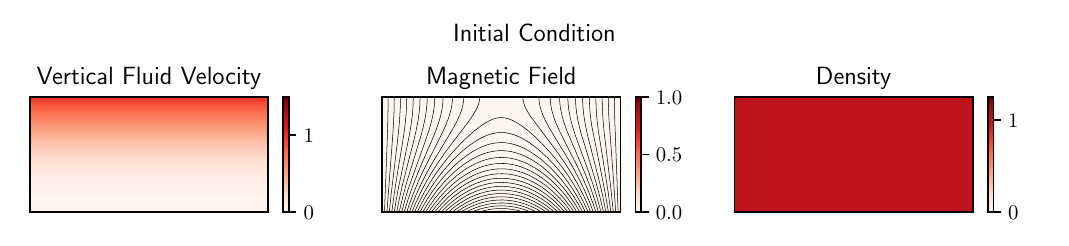}
\plotone{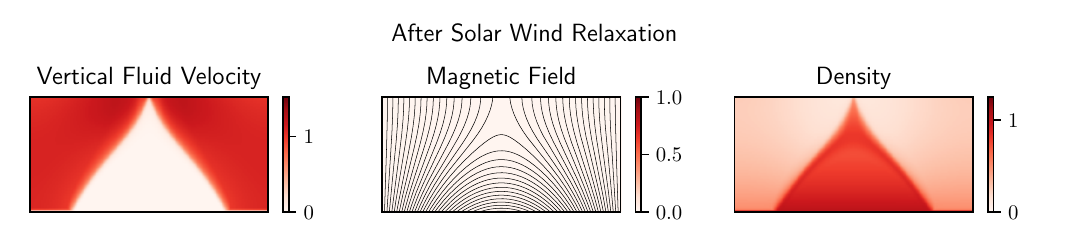}
\caption{Illustration of the initial condition and relaxed `outflow field' state of the MHD simulations. Such an outflow field is obtained in the absence of driven boundary conditions or supergranular diffusion, where the system evolves due to the effect of the solar wind alone. The colormaps show the vertical velocity, out-of-plane magnetic field (zero, in this case) and the fluid density.}
\label{fig:relax}
\end{figure*}

This initial state is far from a steady-state equilibrium, but it does relax to one very quickly relative to the timescale of the magnetic field evolution. The initial conditions and boundary conditions on the velocity (discussed in Section \ref{sec:bconds}) enable the solar wind to be represented self-consistently, as the equilibrium state of the system will naturally have a non-zero vertical fluid velocity. The effect of the solar wind is shown in Figure \ref{fig:relax}, illustrating the initial condition for the MHD simulations (top row) along with the equilibrium state the system reaches in the absence of driven boundary conditions or supergranular diffusion (bottom row).

As the system relaxes to this `outflow field' state, the magnetic field lines at the top of the domain become more radial, with fewer of them connecting back to the surface -- increasing the open flux through the top boundary. Outside the arcade, the vertical fluid velocity is significant and roughly constant. Inside the arcade the fluid is essentially constrained to the magnetic field lines and as such the fluid velocity is comparatively small. During this relaxation period the fluid density within the arcade remains roughly constant, whereas in the region affected by the solar wind the density falls to around a quarter of its initial value, becoming slightly less dense with increasing height. As the system relaxes to this state soon after the simulations are initialised, this can be essentially regarded as the initial condition. A similar process occurs to the magnetic field in the magnetofrictional simulations, although this is due to an imposed solar wind velocity term and the fluid itself is not modelled.

\subsection{Boundary Conditions}
\label{sec:bconds}

\begin{figure*}[ht!]
    \plotone{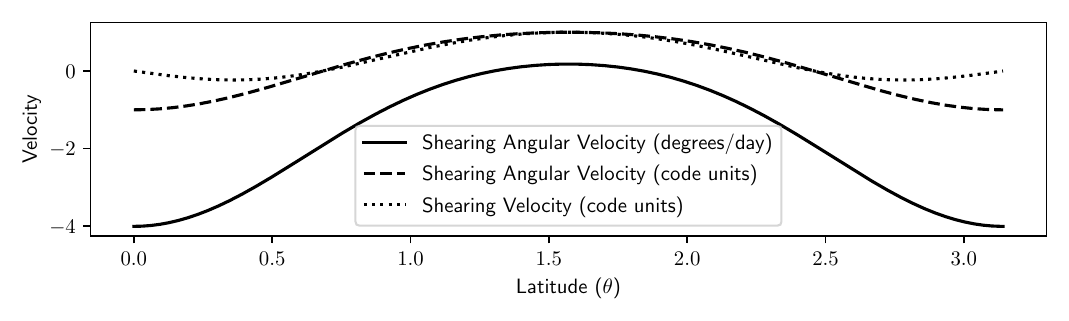}
    \caption{Photospheric shearing rates varying with latitude (measured from the north pole). The angular velocity (solid line) is then shifted into the correct reference frame and to code units (dashed line). It is then multiplied by $\sin{\theta}$ to give a linear velocity on the solar surface (dotted line).}
    \label{fig:shearing}
\end{figure*}

The boundary conditions on the magnetic field are consistent between the three simulation sets. We impose that there is zero perpendicular current on the upper and lower boundaries of the domain, and that the magnetic field is entirely vertical/radial at the side boundaries (or the poles in the axisymmetric simulations).

In the MHD simulations we additionally require boundary conditions on the fluid variables. We impose that the internal energy has zero gradient over the boundary (Neumann boundary conditions), allowing it to evolve to a state independent of its initial value. The fluid density also uses this condition on three of the boundaries, but the density on the lower boundary is held at a fixed value $\rho_0$, imposed at one cell within the domain. This  ensures the density does not fall to zero as the solar wind carries the fluid out the top of the domain.
 
The boundary conditions for the fluid velocity itself are more complex. On the sides of the domain the horizontal ($v_x$) and out-of-plane ($v_z$) velocities are set to zero, preventing flow through the boundaries. There are Neumann boundary conditions on the vertical velocity $v_y$, allowing for upwards flow. Directly on the lower boundary we impose that the vertical velocity is zero, which prevents numerical instabilities. On the top boundary the condition on the vertical fluid velocity is 
\begin{equation}
    v_y(x,1.0) = \textrm{max}( k v_y(x,1.0-\Delta y), 0.0),
\end{equation}
where $1.0-\Delta y$ is the height of the first cell fully within the domain and $k = 1.25$ is a constant that encourages the fluid to accelerate upwards through the top boundary. During arcade eruptions the fluid occasionally changes direction to flow downwards and attempts to suck in material from above the top boundary, which would be numerically problematic. This boundary condition ensures the vertical fluid velocity on the boundary remains non-negative but still allows for these eruptions to occur realistically.

We also impose driven boundary conditions on the lower boundary, representing photospheric shearing and supergranular diffusion. In the magnetofrictional simulations the shearing velocity is added directly to the fictitious velocity field $\mathbf{v}$ on the lower boundary, and in the MHD simulations it is imposed directly on the fluid velocity itself. 

In the axisymmetric simulations we use a realistic profile for the differential rotation rate. As a function of latitude, this is approximately \citep{1983ApJ...270..288S}
\begin{equation}
    V(\theta) = 0.18 - 2.396 \cos^2{\theta} - 1.787 \cos^4(\theta) \,\,\, \rm{degrees}/\rm{day}. \label{eqn:shearing}
\end{equation}
It is common to use the Carrington frame as a reference rotation frame -- rotating at 13.2 degrees/day \citep{2006ApJ...641..577M}. However, in our code we have chosen the reference frame such that the maximum angular velocity in either direction is unity, as illustrated in Figure \ref{fig:shearing}.  This is for consistency with the cartesian simulations where the imposed shearing velocity is simply 
\begin{equation}
V_{\rm {shear}}(x) = V_0 \sin(\pi x).
\end{equation} 
The constant $V_0$ is chosen to be $1.0$ in the magnetofrictional simulations and $0.2$ in MHD. This choice determines the time units used in the code, but otherwise makes no difference to the dynamics.

We also impose an additional magnetic diffusion term on the lower boundary, modelling the effect of unresolved supergranular flows on the photosphere. This diffusion rate ($\eta_0$) is in general much larger than the diffusion rate in the corona ($\eta$), and one of its effects is to bring the footpoints of the magnetic arcade closer together, eventually forming a twisted magnetic flux rope \citep{1989ApJ...343..971V}.

In the magnetofrictional models this diffusion is added as a boundary condition to the electric field $\mathbf{E}$:

\begin{equation}
{\bf E}(x,0) = -\eta_0\frac{\partial B_y(x,0)}{\partial x}{\bf e}_z, 
\end{equation} 
or 
\begin{equation}
{\bf E}(1,\theta) = -\eta_0\frac{1}{r}\frac{\partial B_r(1, \theta)}{\partial \theta}{\bf e}_\phi, 
\end{equation} 
in cartesian and polar coordinates, respectively.

In MHD the photospheric diffusion term cannot be added directly to the electric field in this way. Instead, at each timestep we add an extra diffusive term $\textbf{B}_\textrm{diff}$ to the magnetic field on the lower boundary, defined as
\begin{align}
    B_{x_\textrm{diff}} &= -\eta_0\frac{\Delta t}{\Delta y} \frac{d}{dx}B_y(x,0) \\
    B_{y_\textrm{diff}} &= \eta_0 \Delta t\frac{d^2}{dx^2}B_y(x,0),
\end{align}
where $\Delta t$ is the timestep and $\Delta y$ is the grid resolution in the $y$ direction. This models the photospheric diffusion in MHD by imposing an electric field determined solely from the radial component of the magnetic field on the lower boundary. This results in very similar behaviour to the magnetofrictional equivalent on the lower boundary.

\subsection{Diagnostic Measures}

\label{sec:diags}

In this section we describe the diagnostic measures used to identify eruptive events and ultimately to make predictions of future behavior. We require diagnostics that have a single scalar value which represents the state of the entire system. For the magnetofrictional simulations the diagnostic values will depend only on the strength and configuration of the magnetic field. In our previous paper \citep{2022FrASS...9.9135R} we identify and justify the use of several diagnostic measures. Some of these are also used in this new work, and we also consider some new quantities. The raw diagnostics we calculate are as follows:

\begin{itemize}
\item \textbf{Open Flux}, defined as the sum of the unsigned radial magnetic flux through the upper boundary. An increase in the open flux indicates that fewer magnetic field lines connect back to the surface, and the coronal arcades are stretched upwards, becoming more `open'. We use changes in the open flux to identify arcade eruptions in the magnetofrictional models. 

\item \textbf{Magnetic Energy}, defined as $E_{M} = \int_V\frac{1}{2}B^2\,\mathrm{d}V$. 
For given boundary conditions, the initial potential field has minimal magnetic energy. The energy increases significantly as the field evolves and flux ropes form. There is usually a large decrease in magnetic energy after eruptions. For eruptivity predictions we favour the `Free Magnetic Energy' (see below) rather than the magnetic energy itself.

\item \textbf{Axial Rope Current $I_a$}, defined as the surface integral of the current $\mathbf{j}$ within the rope, in the direction of the rope axis (the $z$ direction in cartesian coordinates or the $\phi$ direction in spherical polar coordinates). In our 2.5D models the `rope' is defined as the region with infinitely-long magnetic field lines that never reach either the photospheric or outer boundaries.

\item \textbf{Poloidal Rope Flux $\Phi_p$} - a measure of the magnetic flux contained within the  rope in the poloidal (in-plane) direction. This is simple to calculate in 2.5D, as it can be defined simply as the flux intersecting a chord between the centre of the rope (where the out-of-plane component of the vector potential $\textbf{A}$ attains its maximum value) and the edge of the domain. 

\item \textbf{Axial Rope Flux $\Phi_a$}, defined as the integral of the magnetic flux in the rope, along the axis (out-of-plane) direction. 

\end{itemize}

For the full MHD simulations we can also measure properties of the fluid. Although ultimately these quantities do not seem to be good predictors of eruptivity, their behavior can be used to identify activity in the system -- for instance, the internal energy peaks during arcade eruptions are more pronounced than any variation in the magnetic field.

\begin{itemize}
\item \textbf{Internal Energy} $\epsilon$, which is here proportional to the temperature of the system, related to the fluid pressure and density by Equation \ref{eqn:gas}.
As a diagnostic we use an integral of this quantity over the entire domain. 

\end{itemize}

In addition to the above, further measurements are obtained by using comparison to a reference potential magnetic field. Such a potential field $\textbf{B}_P$ is defined such that $(\textbf{B} - \textbf{B}_P) \cdot \textbf{n} = 0$ on the domain boundaries, and $\textbf{B}_P = \nabla \Phi$ for some scalar function $\Phi$. This configuration has the lowest-possible magnetic energy for the given boundary conditions. 

In a fully 3D domain the potential field is well-defined, but it is less clear for our 2.5D simulations, as there is no boundary in the third ($z$/$\phi$) dimension. In our previous paper \citep{2022FrASS...9.9135R} we proposed a definition for a reference field in 2.5D space whereby the out-of-plane component of the reference field is a uniform (harmonic) field whose magnetic flux matches the out-of-plane flux of the original magnetic field. This definition has the advantage that an equivalent 3D reference field will converge to it in the limiting case that the domain becomes infinitely long in the out-of-plane dimension. 

In spherical coordinates, we could define the out-of-plane component of the potential field identically to the cartesian case, as the constant average value 
\begin{equation}
    \widetilde{B_{P_\phi}}(r,\theta) = \frac{1}{A} \int B_\phi (r,\theta) \,\mathrm{d}r \,\mathrm{d}\theta, 
\end{equation} 
where $A$ is the cross-sectional area of a hemisphere, equal to $\frac{\pi}{2}(r_1^2 - r_0^2)$. This is simple to calculate and exhibits behavior similar to the cartesian equivalent. However, if such a field is extended to full 3D there would necessarily be a discontinuity at the north and south poles if the average value in each hemisphere is nonzero.

A second approach is to regard the potential field as an axisymmetric field in full 3D. In this case there is no out-of-plane ($\phi$) component at all:
\begin{equation}
    B_{P_\phi}(r,\theta) = 0.
\end{equation} 
The $r$ and $\theta$ components of these potential fields $\textbf{B}_{P}$ and $\widetilde{\textbf{B}_{P}}$ are identical for both definitions, and are calculated independently of the out-of-plane magnetic field. Quantities using both of these definitions and their equivalent vector potentials $\textbf{A}_{P}$ and $\widetilde{\textbf{A}_{P}}$ (calculated by direct integration) are calculated for all three of the models. These `reference-based quantities' are:

\begin{itemize}
    \item \textbf{Relative Helicity $H_R$}
    
    The helicity within a volume $V$ would be defined as $h(V) = \int_V {\bf A} \cdot {\bf B} \, dV, $ where ${\bf A}$ is the vector potential of ${\bf B}$. This quantity is dependent in general on the gauge of ${\bf A}$, and so we use the alternative relative helicity \citep{1984JFM...147..133B}, which is gauge independent:
\begin{equation}
    H_R = \int_V ({\bf A + A}_P) \cdot ({\bf B - B}_P) \,\mathrm{d}V.
\end{equation} 

\item 

\textbf{Current-Carrying Helicity $H_J$} 

Similarly to the relative helicity, this is defined as
\begin{equation}
    H_J = \int_V ({\bf A - A}_P) \cdot ({\bf B - B}_P) \,\mathrm{d}V,
\end{equation}
which is also gauge-independent.

\item \textbf{Free Magnetic Energy $E_F$} 

The free magnetic energy is defined as the magnetic energy of the magnetic field minus the magnetic energy of the reference potential field $\textbf{B}_{P}$,
\begin{equation}
    E_F = \int_V\left(\frac{1}{2}B^2\, - \frac{1}{2}B_P^2\right)\,\mathrm{d}V,
\end{equation}
which is a good indicator of the `excess' energy in the system, as the potential field is the minimum energy state for the given boundary conditions.

\end{itemize}

Each of these three quantities requires construction of a potential field $\textbf{B}_{P}$, which can be calculated using either of the definitions described above. We thus denote quantities calculated using the potential field with no out-of plane component ($\textbf{B}_P$) without a tilde (e.g., $H_R$), and those with an out-of-plane component ($\widetilde{\textbf{B}_P}$) with a tilde (e.g., $\widetilde{E_F}$).

 We previously established \citep{2022FrASS...9.9135R} that no single diagnostic is a good predictor of eruptivity. This is also true in these new simulations and so we instead focus on ratios between them (e.g., $\vert I_a^2/E_F \vert, \vert \widetilde{H_J}/E_F \vert$), chosen with factors such that the ratios are independent of the overall magnetic field strength. 

It must be noted that all of these diagnostic quantities and ratios between them will depend significantly on the size and configuration of the domain, and they are not in general dimensionless. Although understanding these effects will ultimately be necessary, we do not seek to do so in this work and instead will only directly compare diagnostic values where the domain configurations are identical.

\section{Flux Rope Evolution and Eruptions}

\label{sec:behavior}

\subsection{Typical Evolution}

\begin{figure*}[ht!]
\centering
\plotone{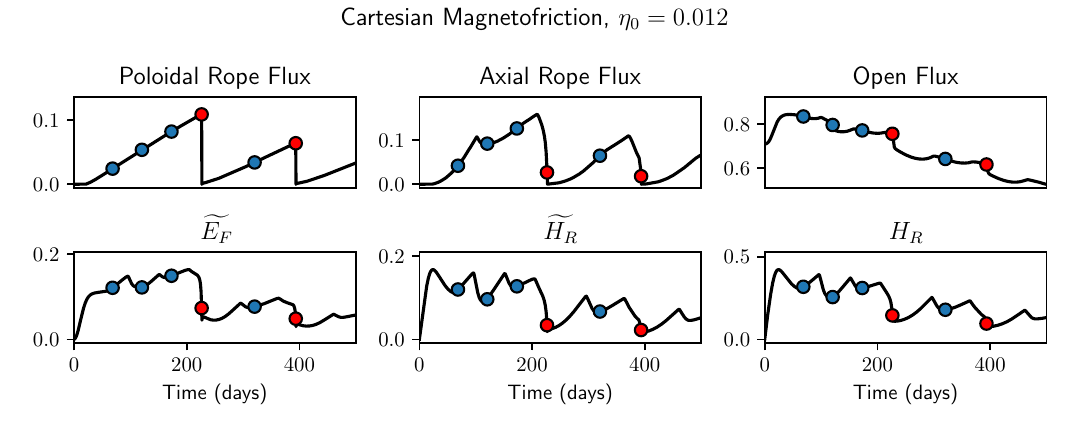}
\plotone{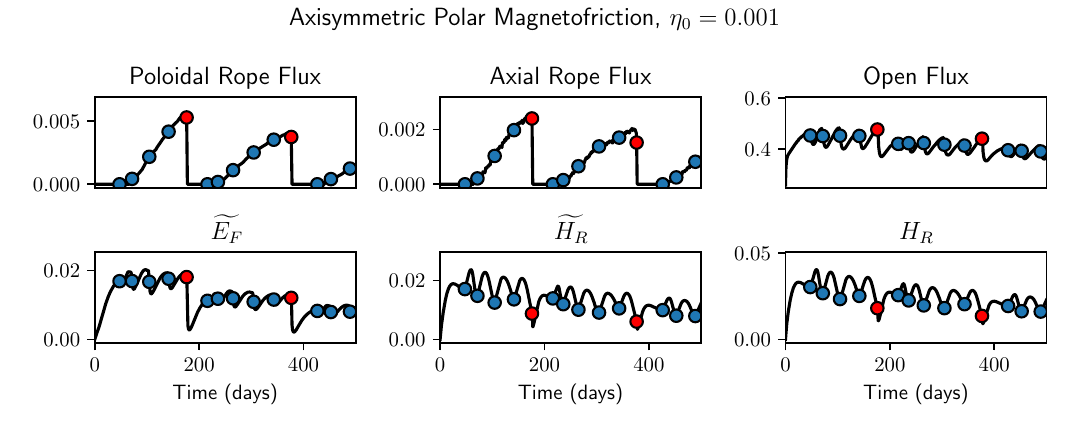}
\plotone{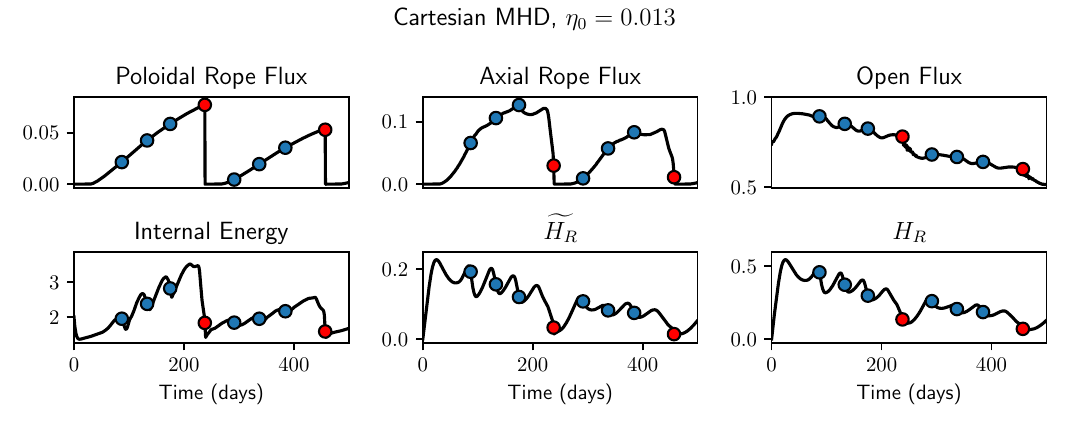}
\caption{Selected diagnostics from a representative run from each of the three simulation sets, in dimensionless units. These units are the same in the two cartesian setups but not in the polar coordinate simulations, due to the differing length scales. Arcade eruptions are represented with blue circles, and flux rope eruptions with red circles. Note that internal energy is plotted for the MHD simulation, whereas free magnetic energy is plotted for the magnetofrictional simulations. The time at which an arcade eruption occurs is given by the time of maximum decrease in either open flux or internal energy, and the time of a flux rope eruption is the point immediately before the poloidal rope flux falls to zero. Values are in general smaller in the axisymmetric case due to small size of the rope compared to the distance unit.}
\label{fig:diagplots}
\end{figure*}

\begin{figure*}[ht!]
    \plotone{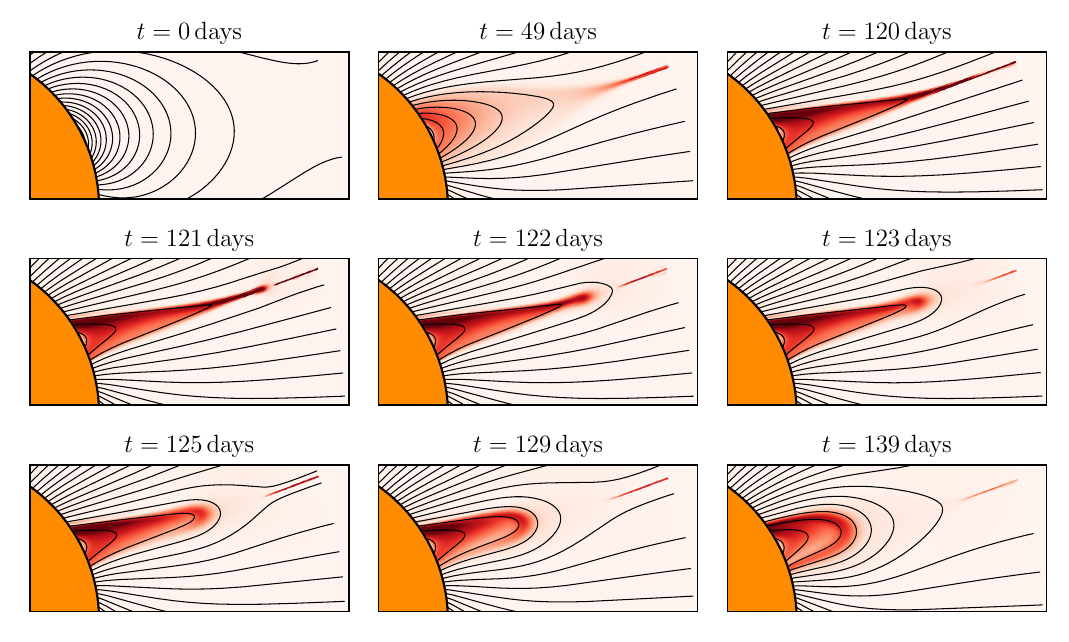}
    \caption{Snapshots from an axisymmetric polar magnetofrictional simulation with no photospheric diffusion ($\eta_0 = 0$, $\nu_0 = 2.0$), showing the shearing of the arcade and formation of a helmet streamer (top panes), followed by an arcade eruption starting at around 120 days. The heatmap represents the out-of-plane current density, and the black lines are the magnetic field lines projected into the plane.}
    \label{fig:arcadeplot}
\end{figure*}

We observe similar behavior in all three sets of simulations. Figure \ref{fig:diagplots} plots selected diagnostics from a single run from each set for comparison. Over the first 10-20 days, the solar wind opens out the magnetic arcade, increasing the open flux through the upper boundary and resulting in a smaller arcade as fewer magnetic field lines loop back to the surface. In the absence of the dynamic lower boundary conditions, the system would remain in equilibrium in this state. Such an `outflow field' is shown in the lower row of Figure \ref{fig:relax}. During this initial phase we also observe significant increases in relative helicity and free energy as the magnetic field evolves further from a potential field. During this period a current sheet develops at the top of the domain (in the out-of-plane direction), forming a helmet streamer \citep{1995ApJ...438L..45L}. Such current sheets are visible in the top snapshots of Figure \ref{fig:arcadeplot} as the arcade opens out.

The differential rotation on the solar surface then causes the magnetic field to become sheared in the out-of-plane direction. Open field lines outside the arcade become sheared near the surface, but as they are not fixed at the top boundary they are free to relax and undo the shear. In the arcade itself the shearing becomes more pronounced, opening out the arcade even further than caused by the effect of the solar wind alone. The free magnetic energy continues to increase  during this period, as does the intensity of the current sheet above the arcade.

After around 50-100 days the energy concentrated in the current sheet becomes too great and there is a rapid loss of equilibrium. There is fast magnetic reconnection at the top of the domain and the arcade quickly moves downwards. We refer to this as an `arcade eruption' \citep{1995ApJ...438L..45L,2021SoPh..296..109B}, indicated by blue circles in Figure \ref{fig:diagplots}. The sequence of such an eruption in an axisymmetric magnetofrictional simulation is shown in Figure \ref{fig:arcadeplot}. In MHD, matter is ejected out of the top of the domain during these eruptions, leading to a large decrease in the internal energy of the system. During these eruptions there are also decreases in open flux (most clearly visible here in the axisymmetric simulations) and free magnetic energy.  

When supergranular diffusion (the parameter $\eta_0$) is significant, we observe the formation and eruption of magnetic flux ropes. The diffusion on the surface causes the magnetic footpoints of the arcade to be  brought together at the polarity inversion line, forming twisted bundles of magnetic flux that no longer connect to the solar surface. Although in reality such a rope would be attached at either end, this is not the case in our 2.5D domains. In the cartesian simulations these ropes are essentially infinitely long in the out-of-plane direction, and in the axisymmetric simulations the ropes wrap entirely around the Sun. 

The presence of the rope in Figure \ref{fig:diagplots} is indicated by nonzero poloidal and axial rope fluxes. In all three sets of simulations the poloidal flux steadily increases until the rope erupts, and is not usually affected by arcade eruptions. The axial flux is affected by arcade eruptions to a greater degree: in the magnetofrictional simulations we observe it steadily increasing, whereas in the MHD simulations it tends to reach a limit or even decrease as the rope evolves. This appears to be the main difference in the dynamics between the two models, and has significant consequences for the prediction of eruptions.

In most of the simulations, a flux rope forms and remains in a semi-equilibrium state for some time. This can be up to several hundred days, but the average time is around 50-100 days after formation. During this period arcade eruptions above the rope can continue to take place. The presence of the rope alters the size and timing of these eruptions, but the qualitative behavior is the same. The rope itself moves downwards during arcade eruptions, and in the MHD simulations this is often followed by a period of damped oscillation as the rope returns to a stable state. Such `kink oscillations' have been observed in the corona, as discussed in \citet{2022ApJ...932L...9K}, where the flux rope fails to erupt in a similar scenario to our `arcade eruptions', although the period of the oscillations in our model is of the order $10,000$ seconds - significantly longer than the observed oscillations.

\subsection{Typical Flux Rope Structure}

\begin{figure*}[ht!]
    \plotone{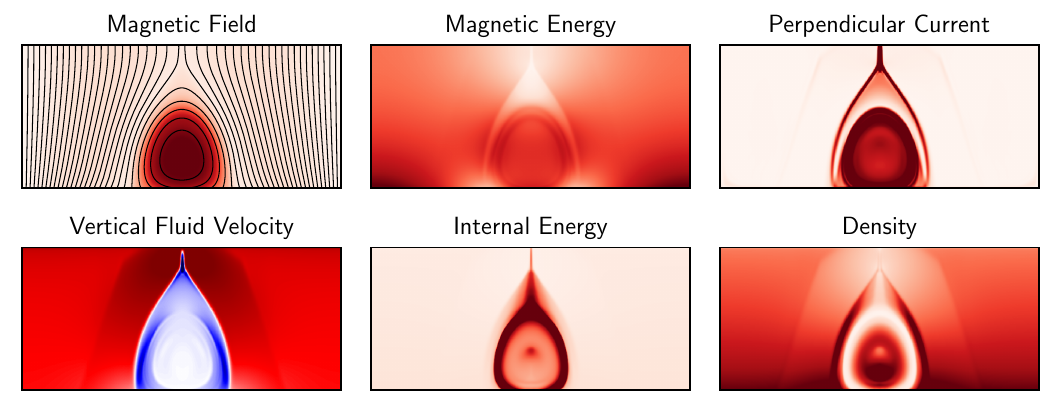}
    \caption{A snapshot of a flux rope from one of the cartesian MHD simulations with $\eta_0 = 0.029$ and $\rho_0 = 1.0$, showing heatmaps of various diagnostic quantities to illustrate its structure. The black lines in the magnetic field plot are the field lines projected onto the plane, and the heatmap represents the out-of-plane component of the field. All quantities except the fluid velocity are strictly positive, increasing from zero (white) to their maximum value (red) on the colormap. The vertical fluid velocity can take negative values within the rope, which are shown in blue.}
    \label{fig:ropesnapshot}
\end{figure*}

The structure of a flux rope in full MHD is displayed in Figure \ref{fig:ropesnapshot}. As expected, we observe that the magnetic field in the rope itself is twisted, with a maximum out-of-plane component of roughly equal magnitude to the radial magnetic field strength through the lower boundary.  As there is comparatively very little magnetic diffusion in the solar corona, the plasma fluid is essentially constrained to move along magnetic field lines (Alfvén's Theorem). We can see evidence of this in the plot of the vertical fluid velocity, which is significant in the region with open magnetic field lines, due to the effect of the solar wind. In the flux rope itself the fluid velocity is comparatively small (less than a quarter of the the velocity outside) or even negative as the fluid must remain within the area of the rope.

The inner `core' of the rope is where the magnetic field is most sheared. In this region the internal energy is relatively low and the fluid density relatively high. The core is surrounded by a region with a considerably lower fluid density (less than 10\% of the core) and much higher temperature, around 5 times that in the core. This region also has a high out-of-plane current as the magnetic field becomes less twisted. This non-uniform shearing at different layers of the flux rope has been noted in both observations and models \citep[see the review by][]{2020RAA....20..165L}. One explanation for this is that layers of flux are added sequentially as the rope forms, and are relatively undisturbed after formation \citep[e.g.][]{2017SoPh..292...25P}. This appears to be the process that we observe here. 

Further out from the centre of the rope there is a region with higher density, similar to that in the core, where the magnetic field itself is not so significantly sheared. Here there is very little perpendicular current, with less than $5\%$ of the maximum attained closer to the centre of the rope. Such a boundary between twisted and untwisted fields has been observed in both analytical models  \citep{1997A&A...325..305D} and MHD simulations \citep{2012A&A...543A.110A}. There is finally a thin current sheet surrounding the whole structure, marking the boundary at which the fluid velocity due to the solar wind becomes significant. We also observe a current sheet above the arcade, as in the cases without photospheric diffusion.

\begin{figure*}[ht!]
    \plotone{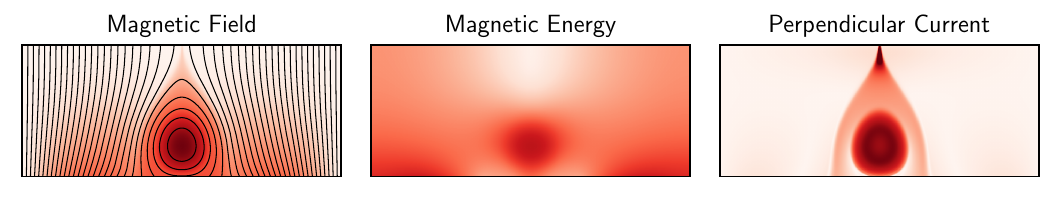}
    \caption{A snapshot of a flux rope in one of the cartesian magnetofrictional simulations with $\eta_0 = 0.005$ and $\nu_0 = 1.0$, showing colormaps of various diagnostic quantities to illustrate its structure. The black lines in the magnetic field plot are the field lines projected onto the plane, and the heatmap represents the out-of-plane component of the field.}
    \label{fig:ropemf}
\end{figure*}

The flux ropes in the magnetofrictional simulations do not have such a complex structure. Such a rope in cartesian coordinates is illustrated for comparison in Figure \ref{fig:ropemf}. Although the overall configuration of the magnetic field is very similar to MHD, the distribution of the current differs in that the current sheet at the top of the domain does not extend down to near the photosphere. This is because the vertical solar wind velocity is applied uniformly across the width of the domain, so there is no shear layer between the open and closed field regions. There is no equivalent of the fluid density or internal energy in magnetofriction, and so these quantities are not comparable.

\begin{figure*}[ht!]
    \plotone{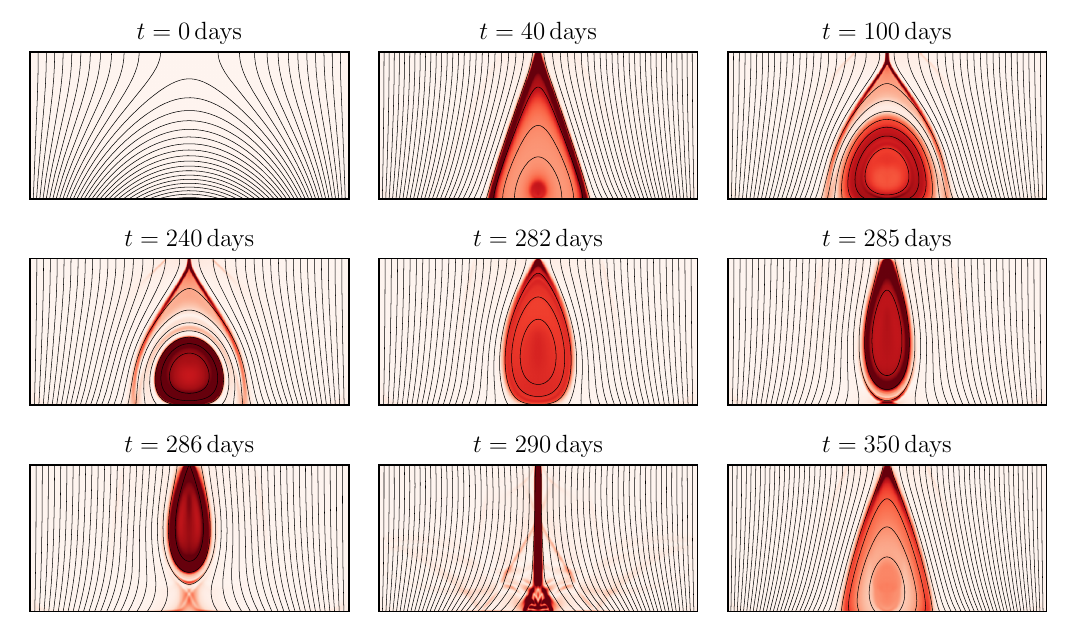}
    \caption{Formation and eruption of a flux rope. This is a cartesian MHD simulation with coronal diffusion $\eta_0 = 0.01$. The heatmaps represent the out-of-plane current density, and the black lines are the magnetic field lines projected into the plane. The rope erupts at around t = 285 days, after which it reforms. The process would then in general repeat.}
    \label{fig:eruption}
\end{figure*}

In most simulations with a flux rope present we observe a catastrophic loss of equilibrium with rapid magnetic reconnection below the rope. The rope itself moves very quickly upward out of the top of the domain -- we call this a `flux rope eruption', indicated by red circles in Figure \ref{fig:diagplots}. During such an eruption there is a significant decrease in all diagnostic quantities, and in the MHD case a large amount of mass is ejected from the domain. The full sequence of flux rope formation and eruption is shown in Figure \ref{fig:eruption}, where both the in-plane magnetic field lines and the out-of-plane current are plotted. 

In most cases, after a flux rope eruption the system still has sufficient energy to continue evolving. If so, after a short period a magnetic arcade reforms and the process restarts. However, the energy in the system is lessened with each eruption and subsequent flux ropes thus have smaller poloidal and axial magnetic fluxes. In general these ropes do not last as long before erupting as those that form initially.

This describes the fundamental processes observed in the simulations. We next discuss the variation in the system behavior based on the system parameters, as well as the differences between cartesian and polar coordinates and between magnetofriction and full MHD.

\subsection{Dependence on Simulation Parameters}

\begin{figure*}[ht!]
\centering
\plotone{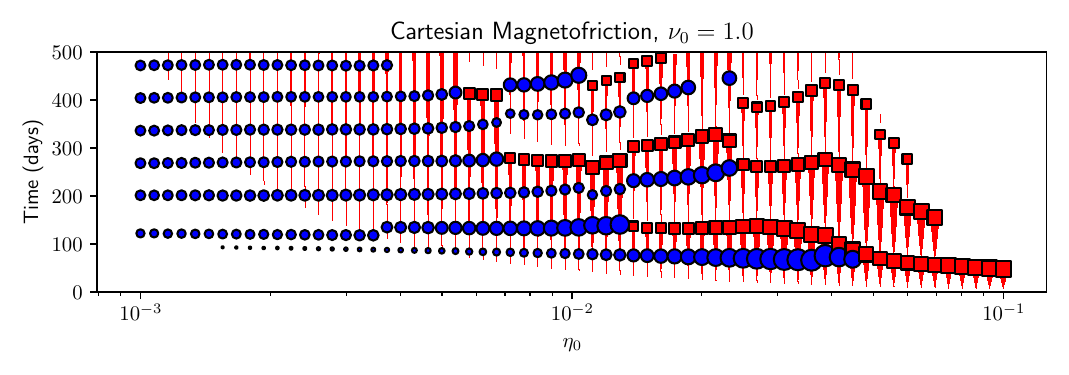}
\plotone{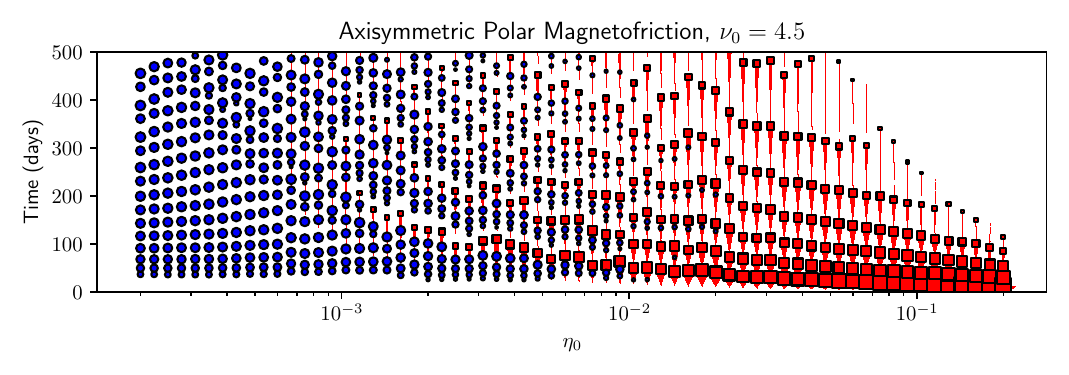}
\plotone{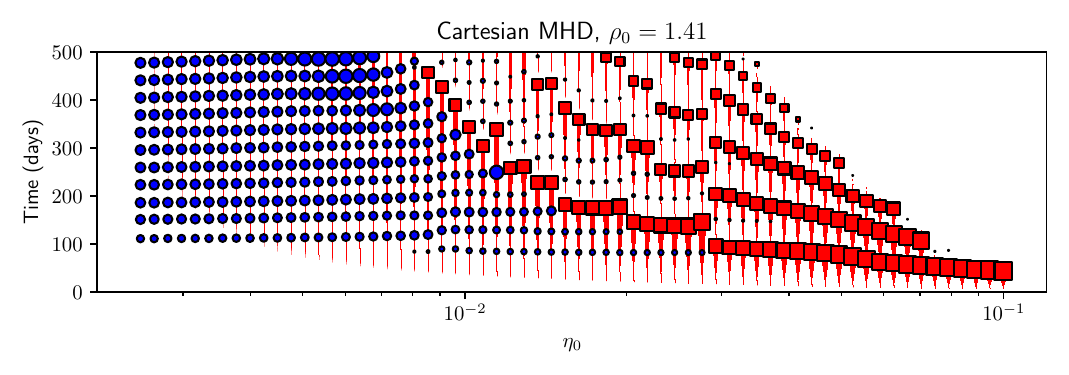}
\caption{Overview of 64 runs from each of the three sets of simulations -- cartesian magnetofriction (top), axisymmetric polar magnetofriction (middle) and cartesian MHD (bottom), varying $\eta_0$ on the horizontal axis. Each simulation corresponds to a vertical line on the plot. The poloidal magnetic flux in the flux rope is represented by the thickness of the red lines. The times of arcade eruptions are represented by blue circles and flux rope eruptions by red squares, respectively. The size of the red squares is proportional to the decrease in poloidal magnetic flux during an eruption, and the size of the blue squares is proportional to the decrease in open flux (magnetofriction) or internal energy (MHD) during each arcade eruption.}
\label{fig:bigplots}
\end{figure*}

To produce a wide array of flux rope behaviors, with differing rope strengths and sizes, we have undertaken a large parameter study within each of the three models. We fix three of the parameters throughout: the outflow velocity factor $v_1$ (set at 50 times the maximum shearing rate), the coronal diffusion $\eta$ ($ = 5 \times 10^{-4}$) and the photospheric shearing rate ($V_0$), as discussed in Section \ref{sec:bconds}.

There are two variable parameters. The most notable is the supergranular diffusion rate $\eta_0$, which is directly related to the rate at which the flux ropes form and erupt. In the magnetofrictional simulations we also vary the magnetofrictional relaxation rate $\nu_0$, whereas in the MHD simulations we vary the initial density $\rho_0$.  

 For each of the three simulation setups we ran 320 simulations in subsets of 64 runs, keeping $\nu_0$ or $\rho_0$ fixed within each subset and varying $\eta_0$. The simulations ran for 500 days, allowing for the formation and eruption of multiple flux ropes in some cases.

The parameter ranges varied based on the simulation setup. For the cartesian magnetofrictional simulations, $10^{-3} < \eta_0 < 10^{-1}$ and $0.5 < \nu_0 < 2$. For the axisymmetric magnetofrictional simulations $2 \times 10^{-4} < \eta_0 < 2 \times 10^{-1}$ and $2 < \nu_0 < 10$. The difference in these ranges is accounted for by the differing unit length scales between the two grid systems. Smaller $\nu_0$ values did not produce realistic behavior as the system takes too long to relax to a force-free state, and for values higher than this the simulations became too computationally intensive (as smaller timesteps are necessary to prevent numerical instability).

In the MHD simulations $2.5 \times 10^{-3} < \eta_0 < 10^{-1}$ (although this is not directly equivalent to the magnetofrictional $\eta_0$ as discussed in Section \ref{sec:bconds}), with the initial density (and the fixed density boundary condition on the lower boundary) in the range $0.5 < \rho_0 < 2.0$. Varying the initial density changes the plasma beta -- the ratio of plasma pressure to magnetic pressure. This does not greatly alter the behavior of the system, but allows us to ensure any eruptivity criteria we later find are independent of the plasma beta.

Figure \ref{fig:bigplots} presents an overview of 64 of the simulations from each of the three models, varying the photospheric diffusion $\eta_0$ on the $x$ axis. The overall pattern is similar in each case. We observe periodic arcade eruptions (the blue circles on the plot) for low $\eta_0$, where there is no (or a very small) flux rope present. For higher values of the photospheric diffusion we observe flux rope formation, indicated by the vertical red lines. There still can be arcade eruptions when a flux rope is present, but when the photospheric diffusion is too high the ropes themselves erupt too quickly to allow for this. The general trend is that for higher $\eta_0$ the ropes form and erupt more quickly. In general ropes reform after erupting, but for very high diffusion rates the system loses enough energy that some ropes do not reform.

Although the overall pattern between the three setups is similar, there are some notable differences. The range of $\eta_0$ at which certain behavior occurs is similar between the two cartesian simulations, but in general ropes form and erupt with smaller $\eta_0$ in the axisymmetric polar simulations, hence the variation in the parameter ranges chosen (the $x$ axes of Figure \ref{fig:bigplots} are adjusted to reflect this). For low $\eta_0$ we observe regular arcade eruptions, but the frequencies are not equivalent in each model -- they are notably less frequent in the cartesian magnetofrictional simulations. The pattern is less regular in the axisymmetric simulations, which is to be expected as the shearing rate is more complex and the domain is larger, allowing the arcade more freedom to move and become asymmetrical (as observed in Figure \ref{fig:arcadeplot}). This freedom also enables the pattern of flux rope eruptions to become less regular than in the cartesian simulations.

In our previous paper it was unclear whether there is a minimum (nonzero) $\eta_0$ below which a flux rope will never erupt. However, we observe for very low $\eta_0$ in cartesian MHD the poloidal rope flux does not necessarily increase monotonically, and indeed in this case the rope will likely slowly diffuse away and not erupt. This diffusion is also observed in some of the cases on the right of Figure \ref{fig:bigplots}, for ropes that have already erupted and reformed. In the axisymmetric simulations ropes do not necessarily form at all for low $\eta_0$. 

A large arcade eruption can prematurely trigger a full flux rope eruption. This phenomenon is clearly observed in the top pane of Figure \ref{fig:bigplots} (the cartesian magnetofrictional simulations). As $\eta_0$ increases, arcade eruptions become larger (indicated by larger blue circles) until they cause the eruption of the rope itself. As the arcade eruptions have very regular frequency, this has resulted in the stepped pattern in the time of first flux rope eruption. Such a pattern is also visible in the cartesian MHD simulations to a lesser extent, but as the arcade eruptions are relatively more frequent it is less clear.

Ultimately, we observe very similar dynamics in all three models, verifying that magnetofriction can accurately represent the qualitative behavior of magnetic flux ropes. We also observe similar behavior between the cartesian and axisymmetric polar simulations, indicating that using cartesian coordinates to model the local dynamics of the corona qualitatively is a valid approach. It remains to discuss the conditions of flux rope instability in each of the models.

\section{Comparison of Scalar Eruptivity Proxies}

\label{sec:skillscores}

\begin{figure*}[ht!]
\centering
\plotone{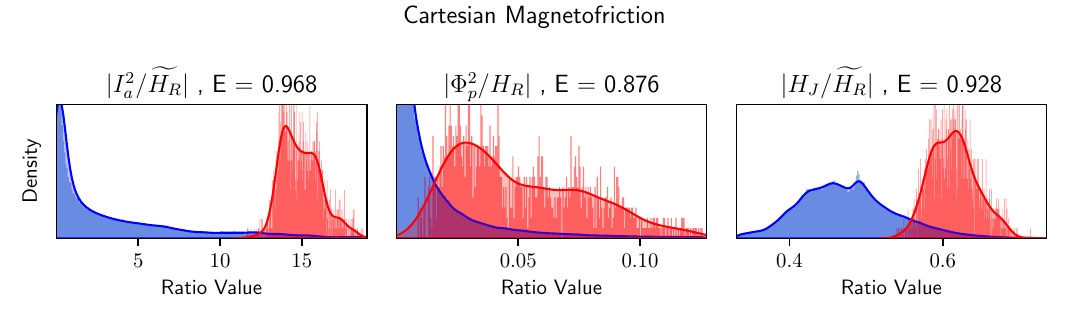}
\plotone{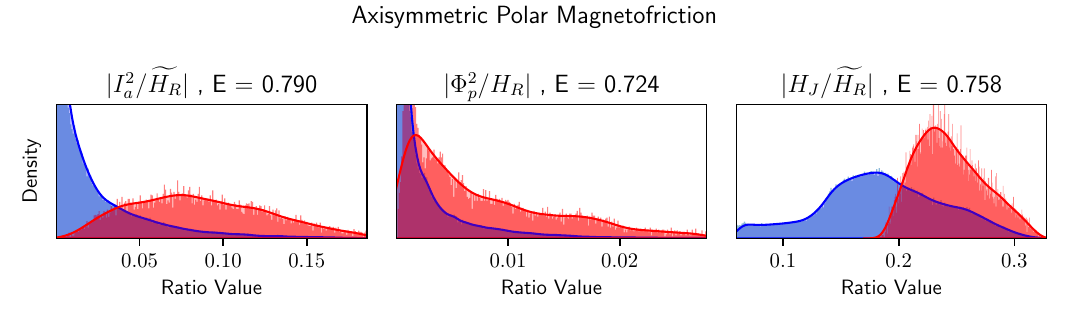}
\plotone{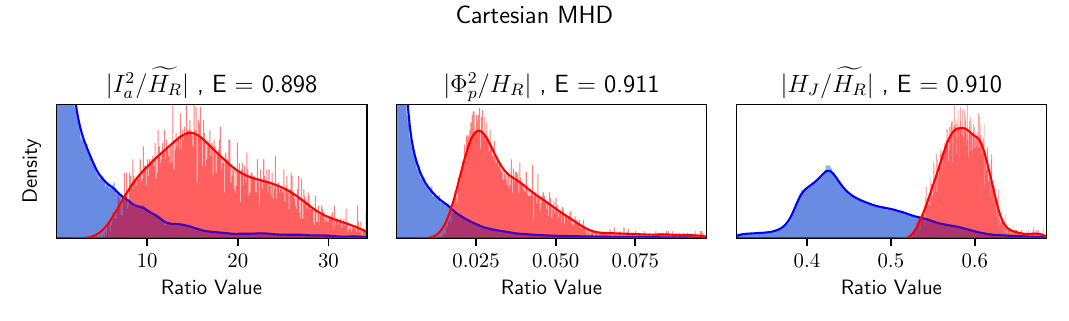}
\caption{Distribution of eruptive and non-eruptive values and the respective distribution curves $R(x,t)$ (red) and $B(x,t)$ (blue) for a flux rope eruption within  $t = 10$ days. The three diagnostic ratios plotted are axial current squared over relative helicity, poloidal flux squared over relative helicity and the current-carrying helicity over relative helicity. Note that the poloidal flux ratio uses the $3D$ definition of the relative helicity (not including the out-of-plane magnetic flux), whereas the others use the $2.5D$ definition. Eruptive points are colored red and non-eruptive points colored blue. The relative heights of each curve give an indication of the likelihood of an eruption occurring within 10 days. The histograms with little red/blue overlap are better predictors.}
\label{fig:histograms}
\end{figure*}

In this section we discuss the possibility of finding a scalar quantity derived from the system diagnostics that can be used to predict an imminent (later) flux rope eruption. Based on our previous magnetofrictional study \citep{2022FrASS...9.9135R} we expect that a single diagnostic value, such as the rope current or open flux, cannot itself be a good predictor of eruptivity due to the large variation in the size and strength of the flux ropes. We found that ratios of the axial rope flux or current to potential field diagnostics (such as the free energy or relative helicity) were good predictors, there being a threshold for each ratio above which an eruption was very likely within a given time. 

\citet{2022FrASS...9.9135R} also found that the `eruptivity index' \citep{2017A&A...601A.125P} -- the ratio of current-carrying helicity to relative helicity -- was a poor predictor of eruptivity, and in fact performed little better than random chance. It should be noted that the definition used in that paper was the `2.5D' version as described in Section \ref{sec:diags} - i.e. $\vert \widetilde{H_J}/\widetilde{H_R} \vert$. Here we also consider the alternative definitions for the reference-based quantities discussed in Section \ref{sec:diags}, and as such there are now four variants of each ratio. 

The diagnostic ratios fall into two categories. The first are ratios of the rope current or flux squared divided by a reference-based quantity (e.g. $\vert \Phi_a ^2 / \widetilde{E_F} \vert$). The second are ratios between two reference-based quantities (e.g. $\vert H_J / \widetilde{H_R} \vert$).

Our procedure for identifying which ratios are good predictors is as follows. For each diagnostic snapshot where a rope exists, the ratio values are sorted according to whether or not they precede a flux rope eruption within a certain time cutoff $t$ (between 10 and 50 days). For each ratio, two histograms are then produced: one for points preceding an eruption and one for points not preceding an eruption. Figure \ref{fig:histograms} shows these histograms in red and blue respectively, for three of the diagnostic ratios. The histograms are normalised to have the same area, resulting in distribution curves $R(x,t)$ and $B(x,t)$ for eruptive and non-eruptive points respectively. This normalization effectively assumes an equal weighting of eruptive and non-eruptive ropes -- this could be improved with prior knowledge of the overall probability of a rope erupting. For instance, regarding the ratio $\vert H_J / \widetilde{H_R} \vert$ at the bottom right of Figure \ref{fig:histograms}, the peak at around 0.6 indicates that if the diagnostics have this value then a flux rope eruption within 10 days is very likely. Good diagnostic predictors will have little overlap between the blue and red regions while bad predictors have a significant overlap.

\begin{figure*}[ht!]
\centering
\plotone{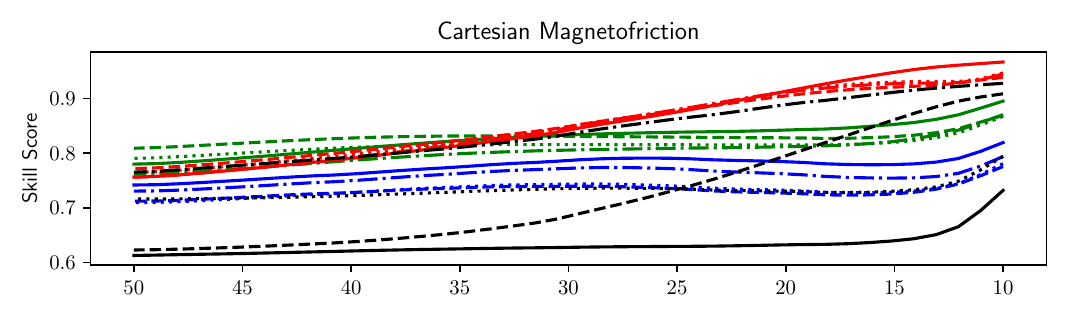}
\plotone{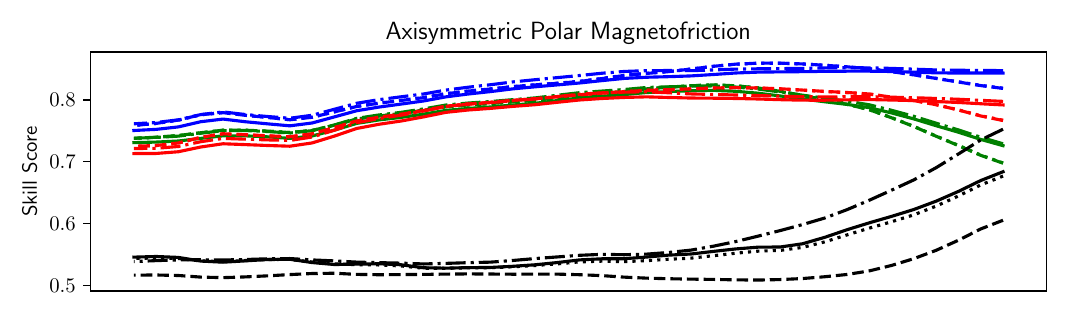}
\plotone{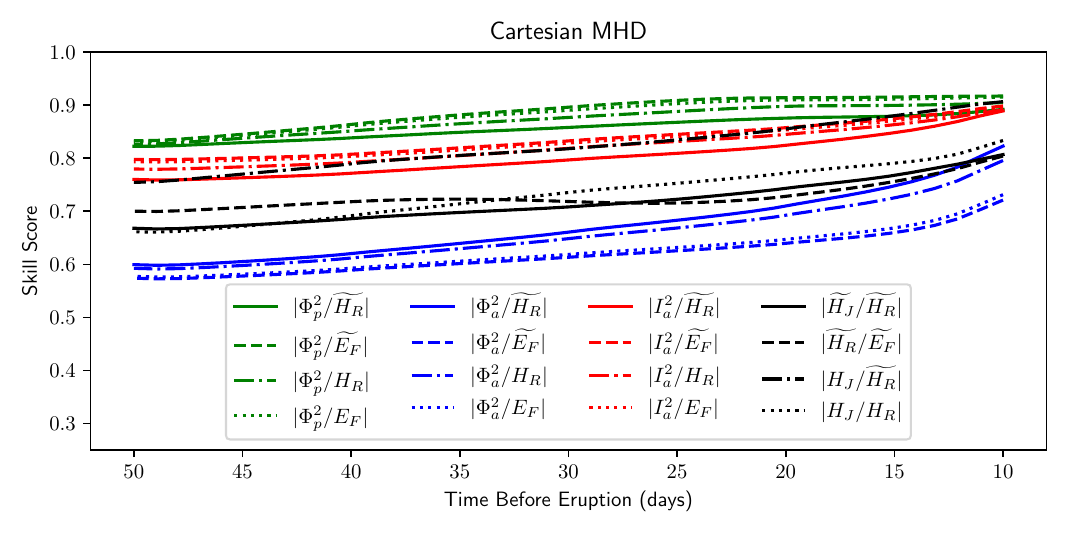}
\caption{The skill scores for selected diagnostic ratios, varying with the time cutoff between a  prediction and the flux rope eruption. Axial rope current and flux are denoted $I_a$ and $\Phi_a$ respectively, poloidal rope flux is denoted $\Phi_p$ and the reference-based quantities are as defined in Section \ref{sec:diags}. The particular diagnostic ratios in black are selected as they perform well in at least one of the simulation sets.}
\label{fig:timevary}
\end{figure*}

We then define the probability of an eruption within time $t$ given a diagnostic ratio value $x$ as
\begin{equation}
    P_e(x,t) = 1 - \left(1 - \frac{R(x,t)}{R(x,t) + B(x,t)} \right) ^2,
\end{equation}
 where the squaring in general slightly increases the predictive ability. We can calculate the accuracy of these probabilities by then comparing against the data, calculating a `skill score'
\begin{equation}
    E(t) = \frac{\Sigma_{\textrm{Erupt } x} \, P_e(x,t) + \Sigma_{\textrm{No Erupt }x}\, (1-P_e(x,t))}{\textrm{Total Number of Values}}.
\end{equation}
If a diagnostic ratio can predict eruptions within a time $t$ perfectly, then it would have skill score $E(t) = 1$ (but this would require complete certainty for each value). If the ratio is no better than random chance than it will have skill score $E=0.5$. The skill scores $E(t)$ for 16 selected diagnostic ratios are shown in Figure \ref{fig:timevary}, for each of the three models. Ratios between reference-based quantities (such as the eruptivity index) are colored black, and the other ratios are colored based on the numerator. 

As in our previous paper, we find that in the cartesian magnetofrictional simulations (top) ratios of the axial current squared (red) to the relative helicity or free energy are the best predictors back to around $t=30$ days before an eruption. In the magnetofrictional simulations these ratios have skill scores up to $E = 0.968$. These ratios also perform well in the MHD simulations, with skill scores around $E = 0.9$. The histograms for one of these ratios ($\lvert I_a^2/\widetilde{H_R}\rvert$) are plotted on the left of Figure \ref{fig:histograms}. We observe that in both of the cartesian simulations (magnetofriction and MHD) the eruptive distribution curve $R(x,10)$ peaks at a value around 15 units, indicating that this criterion for eruptivity is consistent in both magnetofriction and MHD.

Ratios with the poloidal rope flux (green) consistently performed the best in the MHD simulations, with skill scores up to $E=0.911$, although not notably more so than the equivalent ratios with axial current. The ratios using the axial rope flux (blue) as numerator did not perform as well in either scenario.

As expected, predictions of eruptions are more accurate closer to the time of the eruption itself, as seen in the variation in the skill scores plotted in Figure \ref{fig:timevary}. However, the decreased ability to predict eruptions far into the future is counteracted by the general increase in likelihood of an eruption within a larger time frame, explaining why the skill scores do not fall to $E=0.5$ particularly quickly.

The predictive ability of the ratios between two reference-based quantities (such the eruptivity index $\vert H_J/ H_R \rvert$) varies greatly among the three models. The four ratios plotted in Figure \ref{fig:timevary} (in black) are chosen as they are each good predictors in at least one of the models. As we found previously, the eruptivity index calculated using the 2.5D definition $\vert \widetilde{H_J} / \widetilde{H_R} \vert$ (solid black) is not at all a good predictor in the cartesian magnetofrictional simulations. We find it is slightly better in MHD, but has a maximum skill score of only around $0.75$, far less than ratios involving the axial current or poloidal flux.

However, an alternative definition of the eruptivity index - using a mixture of the two reference potential field definitions -- performs very well in both of the cartesian models. This ratio $\vert H_J / \widetilde{H_R} \vert$ is plotted as a dot-dashed black line in Figure \ref{fig:timevary}, and we observe maximum skill scores of $E = 0.928$ in the magnetofrictional simulations and $E = 0.910$ in the MHD simulations. This good predictive ability is also evidenced by the sharp peak at around $0.6$ in the upper and lower right histograms of Figure \ref{fig:histograms}. 

Predictive abilities for all diagnostics are notably lower in the axisymmetric model than the cartesian models. The most basic explanation for this is that all the diagnostic quantities are calculated as integrals over the whole Sun rather than the region immediately surrounding the rope, and so are influenced more heavily by dynamics away from the rope itself. For this reason we suggest that diagnostic measurements calculated by integrating over smaller domains containing just the active region in question are likely to be better predictors.  Ratios with the axial rope flux as numerator (blue) did not perform well in the cartesian simulations, but were (by a small degree) the best predictors in the polar simulations with skill scores of up $E = 0.84$. 

In the axisymmetric simulations, ratios between reference-based quantities do not in general perform well, although the variations on the eruptivity index (dot-dashed, solid and dotted lines in Figure \ref{fig:timevary}) do become better at short timescales, with skill scores up to $E = 0.758$ for the ratio $\vert H_J / \widetilde{H_R} \vert$. Even so, this value is not high enough that it could be reliably used to predict eruptions. This relatively poor performance is likely due not to the different coordinate systems but to the size of the integration domain relative to the rope, and the fact that in this case the current-carrying helicity is capturing irrelevant dynamics from elsewhere in the system. 

\subsection{Consistency between MHD and Magnetofriction}

\begin{table*}[ht!]
\centering
\begin{tabular}{cccccc}
 \hline
 \multicolumn{1}{c}{} & \multicolumn{3}{c}{Peak Eruptive Value} & \multicolumn{2}{c}{Skill Score $E$}  \\
 \hline
 Diagnostic Ratio & MF & MHD & Percentage Difference & MF & MHD\\
 \hline 
$\vert$$\Phi_p^2$/$\widetilde{H_R}$$\vert$ & 0.079 & 0.065 & 22.85\%  & \textbf{0.900} & \textbf{0.893}\\ 
$\vert$$\Phi_p^2$/$\widetilde{E_F}$$\vert$ & 0.058 & 0.040 & 45.32\%  & \textbf{0.873} & \textbf{0.917}\\ 
$\vert$$\Phi_p^2$/$H_R$$\vert$ & 0.025 & 0.025 & \textbf{0.589}\%  & \textbf{0.876} & \textbf{0.905}\\ 
$\vert$$\Phi_p^2$/$E_F$$\vert$ & 0.039 & 0.036 & \textbf{9.168}\%  & \textbf{0.873} & \textbf{0.915}\\  \hline
$\vert$$\Phi_a^2$/$\widetilde{H_R}$$\vert$ & 0.162 & 0.143 & 13.06\%  & 0.825 & 0.830\\ 
$\vert$$\Phi_a^2$/$H_R$$\vert$ & 0.048 & 0.058 & 18.59\%  & 0.800 & 0.803\\  \hline
$\vert$$I_a^2$/$\widetilde{H_R}$$\vert$ & 13.95 & 14.46 & \textbf{3.634}\%  & \textbf{0.968} & \textbf{0.892}\\ 
$\vert$$I_a^2$/$\widetilde{E_F}$$\vert$ & 10.06 & 14.26 & 41.79\%  & \textbf{0.940} & \textbf{0.900}\\ 
$\vert$$I_a^2$/$H_R$$\vert$ & 5.562 & 6.442 & 15.80\%  & \textbf{0.949} & \textbf{0.893}\\ 
$\vert$$I_a^2$/$E_F$$\vert$ & 8.589 & 12.64 & 47.22\%  & \textbf{0.952} & \textbf{0.897}\\  \hline
$\vert$$\widetilde{H_J}$/$\widetilde{H_R}$$\vert$ & 0.261 & 0.278 & \textbf{6.266}\%  & 0.746 & 0.810\\ 
$\vert$$\widetilde{H_R}$/$\widetilde{E_F}$$\vert$ & 0.673 & 0.714 & \textbf{5.975}\%  & \textbf{0.910} & 0.808\\ 
$\vert$$\widetilde{H_R}$/$E_F$$\vert$ & 0.581 & 0.644 & 10.82\%  & 0.758 & 0.817\\ 
$\vert$$H_J$/$\widetilde{H_R}$$\vert$ & 0.617 & 0.585 & \textbf{5.414}\%  & \textbf{0.928} & \textbf{0.908}\\ 
$\vert$$H_J$/$H_R$$\vert$ & 0.224 & 0.240 & \textbf{7.258}\%  & 0.804 & 0.839\\ 
$\vert$$H_R$/$\widetilde{E_F}$$\vert$ & 1.825 & 1.696 & \textbf{7.586}\%  & 0.834 & 0.844\\ 
$\vert$$H_R$/$E_F$$\vert$ & 1.723 & 1.535 & 12.29\%  & \textbf{0.885} & \textbf{0.856}\\  \hline
\end{tabular}

\caption{Table comparing the peak eruptive value and skill scores at time cutoff $t=10$ days for a variety of diagnostic ratios, using data from the cartesian magnetofrictional and MHD simulations. All ratios with skill scores greater than 0.8 in either simulation set are included. Ratios that perform well in a certain aspect are in bold font.}
\label{fig:ratiotable}

\end{table*}

We have shown that certain ratios can perform well as predictors of eruptivity in both magnetofriction and MHD individually, but have not yet considered the ratio values themselves at which an eruption is likely to occur. The domain size, initial conditions and flux rope behavior are directly comparable between cartesian MHD and cartesian magnetofriction, and so we should ideally expect similar peak values for the eruptive histograms in either case.

Table \ref{fig:ratiotable} lists the performance of all diagnostic ratios with skill score greater than 0.8 in either of the cartesian models, using a time cutoff of 10 days. The peak eruptive value (in the first two columns) is the diagnostic value for which an eruption is most likely. Ideally this value would be similar for both the magnetofrictional and MHD simulations. This is true for most of the ratios, especially for those between two reference-based quantities (such as the eruptivity index). The only ratios with large discrepancies (greater than 20\%) are those with the free energy ($E_F$ or $\widetilde{E_F}$) as denominator. 

The fact that these discrepancies exist are an indication that the pre-eruption magnetic fields are not identical in the two models, which is perhaps to be expected. However, the fact that most of the ratios have similar peak eruptive values indicates that magnetofriction could indeed be used as a predictive method for flux rope eruptions of this nature, as long as measuring certain quantities (such as the free energy) is avoided.

Although all of the ratios in Table \ref{fig:ratiotable} with the axial current as numerator had relatively high skill scores in both the magnetofrictional and MHD simulations, only $\vert I_a^2 / \widetilde{H_R}\vert$ has very similar peak eruptive values in both cases -- at around $14$ units. This ratio also has the highest skill score in the magnetofrictional simulations. The equivalent ratio for poloidal flux, but using the other relative helicity definition $\vert \Phi_p^2 / H_R\vert$ also performs well in both models, with values around $0.025$ units indicating an eruption is very likely. 

The ratios between reference-based quantities (black in Figure \ref{fig:timevary}) are in general more consistent between magnetofriction and MHD. This is encouraging, but the skill scores for these ratios are not particularly high compared to those calculated using the rope current or flux. The notable exception to this is the ratio $\vert H_J / \widetilde{H_R}\vert$, which has an excellent skill score in both the MHD and magnetofrictional simulations, and peaks at around 0.6 in both cases.

\section{Discussion}

We have used three independent models (cartesian and polar magnetofriction, and cartesian MHD) to evaluate the ability of a variety of scalar diagnostic quantities to predict the eruption of magnetic flux ropes. We have determined that ratios of the axial rope current squared divided by the relative helicity or free energy are in general the best predictors, but the only one of these ratios that was consistent between magnetofriction and MHD was $\vert I_a^2 / \widetilde{H_R}\vert$, with a peak eruptive value of around $14$ units. Note that such a direct comparison between the models is only valid as the models share the same domain size and setup. The precise threshold value in other simulations will depend on a number of other factors.

Ratios of the poloidal flux squared divided by the relative helicity are also good predictors, but again the only ratio consistent between the two models was $\vert \Phi_p^2 / H_R\vert$, with a peak eruptive value of around $0.025$ units. Note that the ratios between the relative helicity and current-carrying helicity are dimensionless, but the other ratios are not necessarily so and will depend on the chosen length units. 

Ratios with the free energy (defined for either type of reference field) have peak eruptive values that differ significantly between MHD and magnetofriction. This does not necessarily indicate that these ratios cannot be good predictors, but any errors in the extrapolation of the magnetic field may be more likely to alter the respective predictive thresholds. The relative helicity, also defined in either manner, performs better as a denominator and does not have this problem, with (in general) similar peak eruptive values.

Ratios between the relative helicity, current-carrying helicity and free energy were in general more consistent between magnetofriction and MHD. However, although ropes tend to erupt with similar values of these ratios there are significant numbers of ropes that do not, whereas with the axial current ratios there instead tends to be a specific threshold above which ropes are very likely to erupt. 

We have presented two ways to define $H_R, H_J$ and $E_F$, which each have their merits depending on the coordinate system. Of note, the eruptivity index $\vert H_J/H_R\vert$ is not a good predictor of eruptions except when defined as $\vert H_J/\widetilde{H_R} \vert$ -- where the out-of-plane component of the reference potential field is included in the relative helicity $\widetilde{H_R}$ but not in the current-carrying helicity $H_J$. Almost all eruptions in either the cartesian MHD or cartesian magnetofriction simulations occurred when this index had a value around $0.6$ units. 

An explanation for the predictive ability of this particular ratio is that the flux in the rope has more of an effect on the 3D helicity definition than the 2.5D definition. Using the 2.5D definition the out-of-plane component of the magnetic field $\textbf{B}$ is negated by the nonzero component in the potential field $\textbf{B}_P$. Thus the presence and strength of the rope, with its highly out-of-plane magnetic structure, is best quantified using a numerator with the 3D definition (e.g. $H_J$) . In general the ratios with denominators using the 2.5D definition (e.g. $\widetilde{H_R}$) performed better, perhaps as these diagnostics are less affected by the presence of the rope, instead being better indicators of the state of the background magnetic field. 

We note once again that the actual value of any of these ratios will  depend significantly on the size and configuration of the domain (as evidenced by the large difference in these values between the cartesian and axisymmetric simulations). It remains to find a method for establishing the eruptive thresholds for an arbitrarily-sized (or shaped) domain. The errors in the construction of the magnetic field from observed data may also affect the performance of the most successful diagnostics to differing degrees.

We have determined that magnetofriction can be a useful tool for predicting the eruptions of ideally unstable 2.5D flux ropes, where there exist several diagnostic ratios that have consistent thresholds for eruptivity in both magnetofriction and MHD. However, there are many other processes in full MHD (such as breakout or tether cutting) that can be responsible for flux rope eruptions. It remains to be shown whether similar results to ours can be obtained for these cases.

We note that the qualitative behavior of the flux rope system is very similar in either cartesian or polar coordinates, but the diagnostic values behave very differently when the integration domain extends far beyond the influence of the rope. When calculating integrals of helicity in the real corona it would thus be necessary to integrate over a smaller domain just surrounding the rope itself, although in that case care must be taken to choose the location of the boundary of such a domain.

We have shown that the ratio of axial current squared to relative helicity exhibits a threshold above which eruptions are likely. However, in full 3D the rope is less well-defined, and as such the axial current itself would likely be more problematic to measure than the helicity. Thus, when using real observed data it is likely that the eruptivity index $\vert H_J/\widetilde{H_R} \vert$, appropriately defined, may well provide the most accurate indication of an imminent flux rope eruption.


\begin{acknowledgments}
OEKR was supported by a UKRI/STFC PhD studentship, and ARY by UKRI/STFC research grant ST/W00108X/1. The MHD simulations used the LARE2D code from the University of Warwick  (\url{https://github.com/Warwick-Plasma/Lare2d}).

We thank the referee for their comments that have improved the clarity of the paper.
\end{acknowledgments}

\bibliography{main}{}

\begin{thebibliography}{}
\expandafter\ifx\csname natexlab\endcsname\relax\def\natexlab#1{#1}\fi
\providecommand{\url}[1]{\href{#1}{#1}}
\providecommand{\dodoi}[1]{doi:~\href{http://doi.org/#1}{\nolinkurl{#1}}}
\providecommand{\doeprint}[1]{\href{http://ascl.net/#1}{\nolinkurl{http://ascl.net/#1}}}
\providecommand{\doarXiv}[1]{\href{https://arxiv.org/abs/#1}{\nolinkurl{https://arxiv.org/abs/#1}}}

\bibitem[{{Arber} {et~al.}(2001){Arber}, {Longbottom}, {Gerrard}, \&
  {Milne}}]{2001JCoPh.171..151A}
{Arber}, T.~D., {Longbottom}, A.~W., {Gerrard}, C.~L., \& {Milne}, A.~M. 2001,
  Journal of Computational Physics, 171, 151, \dodoi{10.1006/jcph.2001.6780}

\bibitem[{{Aulanier} {et~al.}(2012){Aulanier}, {Janvier}, \&
  {Schmieder}}]{2012A&A...543A.110A}
{Aulanier}, G., {Janvier}, M., \& {Schmieder}, B. 2012, \aap, 543, A110,
  \dodoi{10.1051/0004-6361/201219311}

\bibitem[{{Aulanier} {et~al.}(2010){Aulanier}, {T{\"o}r{\"o}k}, {D{\'e}moulin},
  \& {DeLuca}}]{2010ApJ...708..314A}
{Aulanier}, G., {T{\"o}r{\"o}k}, T., {D{\'e}moulin}, P., \& {DeLuca}, E.~E.
  2010, \apj, 708, 314, \dodoi{10.1088/0004-637X/708/1/314}

\bibitem[{{Berger} \& {Field}(1984)}]{1984JFM...147..133B}
{Berger}, M.~A., \& {Field}, G.~B. 1984, Journal of Fluid Mechanics, 147, 133,
  \dodoi{10.1017/S0022112084002019}

\bibitem[{{Bhowmik} \& {Yeates}(2021)}]{2021SoPh..296..109B}
{Bhowmik}, P., \& {Yeates}, A.~R. 2021, \solphys, 296, 109,
  \dodoi{10.1007/s11207-021-01845-x}

\bibitem[{{Craig} \& {Sneyd}(1986)}]{1986ApJ...311..451C}
{Craig}, I.~J.~D., \& {Sneyd}, A.~D. 1986, \apj, 311, 451,
  \dodoi{10.1086/164785}

\bibitem[{{Demoulin} {et~al.}(1997){Demoulin}, {Bagala}, {Mandrini}, {Henoux},
  \& {Rovira}}]{1997A&A...325..305D}
{Demoulin}, P., {Bagala}, L.~G., {Mandrini}, C.~H., {Henoux}, J.~C., \&
  {Rovira}, M.~G. 1997, \aap, 325, 305

\bibitem[{{Evans} \& {Hawley}(1988)}]{1988ApJ...332..659E}
{Evans}, C.~R., \& {Hawley}, J.~F. 1988, \apj, 332, 659, \dodoi{10.1086/166684}

\bibitem[{{Forbes} {et~al.}(2006){Forbes}, {Linker}, {Chen}, {Cid}, {K{\'o}ta},
  {Lee}, {Mann}, {Miki{\'c}}, {Potgieter}, {Schmidt}, {Siscoe}, {Vainio},
  {Antiochos}, \& {Riley}}]{2006SSRv..123..251F}
{Forbes}, T.~G., {Linker}, J.~A., {Chen}, J., {et~al.} 2006, Space Science
  Reviews, 123, 251, \dodoi{10.1007/s11214-006-9019-8}

\bibitem[{{Gupta} {et~al.}(2021){Gupta}, {Thalmann}, \&
  {Veronig}}]{2021A&A...653A..69G}
{Gupta}, M., {Thalmann}, J.~K., \& {Veronig}, A.~M. 2021, \aap, 653, A69,
  \dodoi{10.1051/0004-6361/202140591}

\bibitem[{{Hoeksema} {et~al.}(2020){Hoeksema}, {Abbett}, {Bercik}, {Cheung},
  {DeRosa}, {Fisher}, {Hayashi}, {Kazachenko}, {Liu}, {Lumme}, {Lynch}, {Sun},
  \& {Welsch}}]{2020ApJS..250...28H}
{Hoeksema}, J.~T., {Abbett}, W.~P., {Bercik}, D.~J., {et~al.} 2020, \apjs, 250,
  28, \dodoi{10.3847/1538-4365/abb3fb}

\bibitem[{{Kliem} \& {T{\"o}r{\"o}k}(2006)}]{2006PhRvL..96y5002K}
{Kliem}, B., \& {T{\"o}r{\"o}k}, T. 2006, Physical Review Letters, 96, 255002,
  \dodoi{10.1103/PhysRevLett.96.255002}

\bibitem[{{Kumar} {et~al.}(2022){Kumar}, {Nakariakov}, {Karpen}, {Richard
  DeVore}, \& {Cho}}]{2022ApJ...932L...9K}
{Kumar}, P., {Nakariakov}, V.~M., {Karpen}, J.~T., {Richard DeVore}, C., \&
  {Cho}, K.-S. 2022, \apjl, 932, L9, \dodoi{10.3847/2041-8213/ac6e3e}

\bibitem[{{Kuperus} \& {Raadu}(1974)}]{1974A&A....31..189K}
{Kuperus}, M., \& {Raadu}, M.~A. 1974, Astronomy \& Astrophysics, 31, 189

\bibitem[{{Leake} {et~al.}(2014){Leake}, {Linton}, \&
  {Antiochos}}]{2014ApJ...787...46L}
{Leake}, J.~E., {Linton}, M.~G., \& {Antiochos}, S.~K. 2014, \apj, 787, 46,
  \dodoi{10.1088/0004-637X/787/1/46}

\bibitem[{{Leake} {et~al.}(2013){Leake}, {Linton}, \&
  {T{\"o}r{\"o}k}}]{2013ApJ...778...99L}
{Leake}, J.~E., {Linton}, M.~G., \& {T{\"o}r{\"o}k}, T. 2013, \apj, 778, 99,
  \dodoi{10.1088/0004-637X/778/2/99}

\bibitem[{{Linker} \& {Mikic}(1995)}]{1995ApJ...438L..45L}
{Linker}, J.~A., \& {Mikic}, Z. 1995, The Astrophysical Journal, 438, L45,
  \dodoi{10.1086/187711}

\bibitem[{{Liu}(2020)}]{2020RAA....20..165L}
{Liu}, R. 2020, Research in Astronomy and Astrophysics, 20, 165,
  \dodoi{10.1088/1674-4527/20/10/165}

\bibitem[{{Mackay} \& {van Ballegooijen}(2006)}]{2006ApJ...641..577M}
{Mackay}, D.~H., \& {van Ballegooijen}, A.~A. 2006, \apj, 641, 577,
  \dodoi{10.1086/500425}

\bibitem[{{Mackay} \& {Yeates}(2012)}]{2012LRSP....9....6M}
{Mackay}, D.~H., \& {Yeates}, A.~R. 2012, Living Reviews in Solar Physics, 9,
  6, \dodoi{10.12942/lrsp-2012-6}

\bibitem[{{Pagano} {et~al.}(2013){Pagano}, {Mackay}, \&
  {Poedts}}]{2013A&A...554A..77P}
{Pagano}, P., {Mackay}, D.~H., \& {Poedts}, S. 2013, \aap, 554, A77,
  \dodoi{10.1051/0004-6361/201220947}

\bibitem[{{Pariat} {et~al.}(2017){Pariat}, {Leake}, {Valori}, {Linton},
  {Zuccarello}, \& {Dalmasse}}]{2017A&A...601A.125P}
{Pariat}, E., {Leake}, J.~E., {Valori}, G., {et~al.} 2017, Astronomy \&
  Astrophysics, 601, A125, \dodoi{10.1051/0004-6361/201630043}

\bibitem[{{Pariat} {et~al.}(2023){Pariat}, {Wyper}, \&
  {Linan}}]{2023A&A...669A..33P}
{Pariat}, E., {Wyper}, P.~F., \& {Linan}, L. 2023, \aap, 669, A33,
  \dodoi{10.1051/0004-6361/202245142}

\bibitem[{{Parker}(1958)}]{1958ApJ...128..664P}
{Parker}, E.~N. 1958, \apj, 128, 664, \dodoi{10.1086/146579}

\bibitem[{{Priest} \& {Longcope}(2017)}]{2017SoPh..292...25P}
{Priest}, E.~R., \& {Longcope}, D.~W. 2017, \solphys, 292, 25,
  \dodoi{10.1007/s11207-016-1049-0}

\bibitem[{{Rice} \& {Yeates}(2021)}]{2021ApJ...923...57R}
{Rice}, O. E.~K., \& {Yeates}, A.~R. 2021, \apj, 923, 57,
  \dodoi{10.3847/1538-4357/ac2c71}

\bibitem[{{Rice} \& {Yeates}(2022)}]{2022FrASS...9.9135R}
---. 2022, Frontiers in Astronomy and Space Sciences, 9, 849135,
  \dodoi{10.3389/fspas.2022.849135}

\bibitem[{{Snodgrass}(1983)}]{1983ApJ...270..288S}
{Snodgrass}, H.~B. 1983, The Astrophysical Journal, 270, 288,
  \dodoi{10.1086/161121}

\bibitem[{{Svalgaard} \& {Wilcox}(1978)}]{1978ARA&A..16..429S}
{Svalgaard}, L., \& {Wilcox}, J.~M. 1978, \araa, 16, 429,
  \dodoi{10.1146/annurev.aa.16.090178.002241}

\bibitem[{{van Ballegooijen} \& {Martens}(1989)}]{1989ApJ...343..971V}
{van Ballegooijen}, A.~A., \& {Martens}, P.~C.~H. 1989, The Astrophysical
  Journal, 343, 971, \dodoi{10.1086/167766}

\bibitem[{{Wang} {et~al.}(2005){Wang}, {Lean}, \&
  {Sheeley}}]{2005ApJ...625..522W}
{Wang}, Y.~M., {Lean}, J.~L., \& {Sheeley}, N.~R., J. 2005, \apj, 625, 522,
  \dodoi{10.1086/429689}

\bibitem[{{Webb} \& {Howard}(2012)}]{2012LRSP....9....3W}
{Webb}, D.~F., \& {Howard}, T.~A. 2012, Living Reviews in Solar Physics, 9, 3,
  \dodoi{10.12942/lrsp-2012-3}

\bibitem[{{Wiegelmann} \& {Sakurai}(2012)}]{2012LRSP....9....5W}
{Wiegelmann}, T., \& {Sakurai}, T. 2012, Living Reviews in Solar Physics, 9, 5,
  \dodoi{10.12942/lrsp-2012-5}

\bibitem[{{Yang} {et~al.}(1986){Yang}, {Sturrock}, \&
  {Antiochos}}]{1986ApJ...309..383Y}
{Yang}, W.~H., {Sturrock}, P.~A., \& {Antiochos}, S.~K. 1986, The Astrophysical
  Journal, 309, 383, \dodoi{10.1086/164610}

\bibitem[{{Yeates}(2014)}]{2014SoPh..289..631Y}
{Yeates}, A.~R. 2014, \solphys, 289, 631, \dodoi{10.1007/s11207-013-0301-0}

\bibitem[{{Yee}(1966)}]{1966ITAP...14..302Y}
{Yee}, K. 1966, IEEE Transactions on Antennas and Propagation, 14, 302,
  \dodoi{10.1109/TAP.1966.1138693}

\bibitem[{{Zuccarello} {et~al.}(2015){Zuccarello}, {Aulanier}, \&
  {Gilchrist}}]{2015ApJ...814..126Z}
{Zuccarello}, F.~P., {Aulanier}, G., \& {Gilchrist}, S.~A. 2015, \apj, 814,
  126, \dodoi{10.1088/0004-637X/814/2/126}

\bibitem[{{Zuccarello} {et~al.}(2018){Zuccarello}, {Pariat}, {Valori}, \&
  {Linan}}]{2018ApJ...863...41Z}
{Zuccarello}, F.~P., {Pariat}, E., {Valori}, G., \& {Linan}, L. 2018, \apj,
  863, 41, \dodoi{10.3847/1538-4357/aacdfc}

\end{thebibliography}
\bibliographystyle{aasjournal}



\end{document}